\documentclass[12pt]{article}
\usepackage{amsmath,amssymb,epsfig}

\def\nonu{\nonumber}
\newcommand{\tr}{\mbox{Tr}}
\newcommand{\tQ}{\widetilde{Q}}

\newcommand{\ra}{\rightarrow}

\newcommand{\be}{\begin{equation}}
\newcommand{\ee}{\end{equation}}
\newcommand{\ba}{\begin{eqnarray}}
\newcommand{\ea}{\end{eqnarray}}
\newcommand{\bi}{\begin{itemize}}
\newcommand{\ei}{\end{itemize}}
\newcommand{\Tr}{{\rm Tr}}

\newcommand{\p}{\partial}

\newcommand{\Ncal}{{\mathcal N}}
\newcommand{\Mcal}{{\mathcal M}}

\newcommand{\nn}{\nonumber}
\renewcommand{\bar}{\overline}
\renewcommand{\tilde}{\widetilde}
\renewcommand{\hat}{\widehat}
\begin{document}

\thispagestyle{empty} \addtocounter{page}{-1}
\begin{flushright}
CALT-68-2574 \\
UT-05-11 \\
{\tt hep-th/0508189}\\
\end{flushright}

\vspace*{1.8cm} \centerline{ \Large \bf  Higgsing and Superpotential
Deformations of } \vspace{0.5cm} \centerline{\Large \bf  ADE
Superconformal Theories} \vskip0.3cm
%1. Deformations of the ADE superconformal field theories.
%2. Testing the $a$-theorem conjecture for Higgsing and mesonic superpotential deformations
%3. Further study of four dimensional superconformal theories
%4.
%(``Mesonic'' and ``Field'' removed)

\vspace*{2.0cm} \centerline{{\bf Takuya Okuda}$^1$ \ and  \ {\bf
Yutaka Ookouchi}$^{2}$} \vspace*{1.0cm} \centerline{\it $^1$
California Institute of Technology, Pasadena , CA 91125, USA}
\vspace*{0.2cm} \centerline{\it $^2$Department of Physics,
 University of Tokyo, Tokyo 113-0033, Japan} \vspace*{0.8cm} \centerline{\tt
takuya@theory.caltech.edu, \qquad
ookouchi@hep-th.phys.s.u-tokyo.ac.jp} \vskip2cm

\vspace{1cm} \centerline{\bf Abstract} \vspace*{0.5cm}

We study large classes of renormalization group flows, driven by
scalar expectation values or mesonic superpotential terms, away from
the conformal fixed points of the $4d$ supersymmetric gauge theories
with $ADE$-type superpotentials. The $a$-maximization procedure
allows us to compute the $R$ charges and to check the $a$-theorem
conjecture. For a theory obtained by Higgsing the $D_{k+2}$ theory,
we use the magnetic dual description proposed by Brodie to determine
the parameter region where the resulting theory is at a non-trivial
conformal fixed point.

\baselineskip=18pt
\newpage
\renewcommand{\theequation}{\arabic{section}\mbox{.}\arabic{equation}}

%%%%%%%%%%%%%%%%%%%%%%%%%%%%%%%%%%%%%%%%%%%%%%%%%%%%%%%%%%%%%%%%%%%%%%%%%%%%%
\section{Introduction}\label{intro}
\setcounter{equation}{0}
%%%%%%%%%%%%%%%%%%%%%%%%%%%%%%%%%%%%%%%%%%%%%%%%%%%%%%%%%%%%%%%%%%%%%%%%%%%%%

The $a$-maximization procedure, as shown by Intriligator and Wecht
in \cite{IW}, determines the scaling dimensions of chiral primary
operators and allows one to study superconformal fixed points in
four dimensional supersymmetric gauge theories. When the
superconformal theory admits flavor symmetries (the global
symmetries that commute with supercharges), the $R$ symmetry can mix
with such symmetries and is not unique as far as the supersymmetry
algebra is concerned. However, the $R$ symmetry that appears in the
superconformal algebra $SU(2|2,1)$ is unique. The unique
superconformal $R$ symmetry is the one that maximizes the ``trial
$a$-function'' among all the possible $R$ symmetries.
Recently other criterions to fix the ambiguities were proposed in \cite{MSY,Tachikawa,New1,New2}.
The superconformal $R$ charge of a chiral primary operator determined in
this way is proportional to its scaling dimension.
The knowledge of the superconformal $R$ charges also allows one to check the
conjectural ``$a$-theorem''\cite{Cardy}, which states that the value
of $a$ decreases under any renormalizaton group flow. If the theory
contains accidental symmetries that are not apparent from the
underlying Lagrangian, $a$-maximization does not always allow one
to determine the superconformal $R$ symmetry. One exception is the
case where some operators violate the unitarity bound in \cite{KPS}.
According to their proposal, when the naive $a$-maximization
predicts the scaling dimension a chiral operator to be below the
minimum value allowed by unitarity, the operator becomes free and is
decoupled from the rest of the interacting theory. The decoupling of
the operator gives rise to an accidental $U(1)$ symmetry acting on
the free field. By assuming that the theory possesses no other
accidental symmetries, the validity of the conjectured $a$-theorem
was confirmed in \cite{IW,KPS,IWADE,BIWW,Kutasov} in such a
situation.

By assuming that a theory has no accidental symmetries other than
those associated with the operators that hit the unitarity bound, it
is possible to explore a variety of RG flows among superconformal
gauge theories, testing the $a$-theorem conjecture for each flow. In
\cite{IWADE}, Intriligator and Wecht studied the RG fixed points
obtained from the two-adjoint SQCD, that is, the supersymmetric
$SU(N)$ gauge theory with $N_f$ pairs of fundamental and
anti-fundamental chiral fields $(Q^i,\widetilde{Q}_{i})$ and
two adjoint chiral fields $X$ and $Y$. They found that, under the
assumption that the superpotential is the sum of single trace
operators made from adjoint fields, the fixed points fall into the
$ADE$ classification, just like the classification of simple
singularities \cite{Arnold}. To date, the reason why these fixed points are
classified by the $ADE$ Dynkin diagrams remains mysterious. The
symbols for the fixed points and the corresponding superpotentials
are the following:
\begin{eqnarray}
&&\widehat{O}: W=0,~~ ~\widehat{A}: W=\tr Y^2,~~~ \widehat{D}: W=\tr XY^2,~~~\widehat{E}: W=\tr Y^3,\nonumber\\
&&A_k: W=\tr(X^{k+1}+Y^2),~~~D_{k+2}:W=\tr(X^{k+1}+XY^2),\label{ADEpotentials}\\
&&E_6: W=\tr(Y^3+X^4), ~~~E_7: W=\tr(Y^3+YX^3),~~~E_8:
W=\tr(Y^3+X^5).\nonumber
\end{eqnarray}
The hatted symbols represent the parent theories from which the
indexed theories are obtained by single-trace superpotential
deformations. The $A$-type theories ($\widehat{A}$ and $A_{k+1}$)
can be considered as one-adjoint SQCD since $Y$ is massive and
gets integrated out in the IR.
In eq.(\ref{ADEpotentials}), as we will throughout the paper,
we omitted the coefficients in front of operators
in the superpotential.

The aim of this paper is to extend the class of conformal fixed
points of the two-adjoint theory obtained through the RG flows
driven by scalar expectation values (Higgsing), as well as
deformations by multi-trace and mesonic superpotential terms. We
now explain motivations for this investigation.

$a$-maximization itself is closely related to the $a$-theorem
conjecture \cite{IW}. Since the introduction of a new interaction
generically eliminates some of the flavor symmetries, the maximum
value of $a$ tends to be smaller at the IR fixed point than at the
UV fixed point. This is not a proof because of two loopholes: i)
There can be accidental symmetries, at the IR fixed point, which
enlarge the parameter space where $a$ is to be maximized . ii) The
value of $a$ obtained by $a$-maximization is only a local maximum.
The second loophole was closed by Kutasov in \cite{Kutasov}, who
provided a way to understand the $a$-theorem for RG flows caused by a
deformation of the superpotential or a gauge interaction. His idea
was to extend $a$ to a function of certain parameters (Lagrange
multipliers enforcing the vanishing of the beta functions)
interpolating fixed points. The Lagrange multipliers were then
identified with the coupling constants \cite{Kutasov,KS}. While the RG flows
driven by a superpotential deformation or a gauge interaction are
expected to obey the $a$-theorem as reviewed above, Higgsing RG flows
are still to be better understood. In this paper we will consider
several classes of Higgsing RG flows and check the validity of the
$a$-theorem.

Although the Lagrange multiplier method is expected to work
for RG flows triggered by superpotential deformations,
there is no general proof that the $a$-theorem is satisfied
even in these cases.
This provides us with the first motivation for
studying deformations of the $ADE$
conformal theories by mesonic superpotential terms,
which supply us with many examples of RG flows where we can test
the $a$-theorem.
The second motivation is that two of
the mesonic terms, namely $\widetilde{Q}_i X Q^i$ and
$\widetilde{Q}_i Y Q^i$, naturally arise when one attempts to
construct the gauge theories with the $ADE$ superpotentials via
D-branes in a non-trivial geometry.\footnote{In \cite{CKV},
two-adjoint ${\mathcal N}=1$ theories without quarks were engineered
by non-compact D-branes. Tuning the geometry allows one to have
$ADE$ superpotentials. One can further incorporate quark fields by
introducing non-compact D-branes, at the cost of adding
$\widetilde{Q}_i X Q^i$ and $\widetilde{Q}_i Y Q^i$ in the
superpotential.} Unfortunately, our results will show that the $ADE$
conformal theories do not possess mesonic operators that are
relevant. Thus the theories realized by D-branes are not connected
to the $ADE$ conformal theories by RG flows.

The inclusion of multi-trace operators gives rise to a manifold of
fixed points. This can be seen as follows. A conformal fixed point
is where the beta functions for all the couplings simultaneously
vanish. For a generic theory, since there are as many equations as
the variables, a conformal fixed point occurs at an isolated point
in the space of couplings. As discussed in \cite{LS}, a manifold of
fixed points is realized when not all the beta functions are
functionally independent. This is clearly the case when the
superpotential has more than one operator consisting of the same
number of elementary chiral fields. For example, $\tr X^5$, which
gives rise to the $A_4$ theory, and $\tr X^2 \tr X^3$ have identical
beta functions because they depend on the couplings only through the
anomalous dimension of the chiral field $X$. This implies that each
of the $A_k$, $D_{k+2}$ and $E_{6,7,8}$ theories is part of a
continuous manifold of fixed points. Similarly, more complicated
operators containing mesons, such as $(\widetilde{Q}_j X^m Y^n Q^i)
(\widetilde{Q}_i Y^s Q^j)$ and $(\widetilde{Q}_j X^m Y^n Q^j)(
\widetilde{Q}_i Y^s Q^i)$ with
 differently contracted indices,
 also give rise to a manifold of fixed points.

Finding new conformal fixed points is by itself a worthwhile
objective.
Because we encounter so many RG flows that we can study,
we usually focus on new interacting conformal field theories.
For example, when we find that an RG flow leads to a product of known
conformal theories without any interaction between them,
we do not always proceed further
though it is certainly possible to compare the values
of the central charges and check the $a$-theorem conjecture.

Throughout the paper, we will work in the large $N$ limit
\begin{eqnarray}
N_i\rightarrow \infty,\quad N_f\rightarrow\infty,\quad {N\over
N_f}{~\rm and}~{N_{i=1,2} \over N_f} \quad {\rm fixed },
\label{largeN}
\end{eqnarray}
as in \cite{IWADE} and related papers. All the calculations can in
principle be done exactly, but the expressions simplify considerably in
the large $N$ approximation. It is usually convenient to define the
central charge $a$ in terms of the 't Hooft anomalies as $3\tr
R^3-\tr R$. This convention differs from the earlier literature by
the overall normalization that is irrelevant for our purposes. In
fact throughout our paper we use the definition
\begin{eqnarray}
a\equiv {1\over N_f^2} \left(3\tr R^3-\tr R \right),
\label{aconvention}
\end{eqnarray}
which is even more convenient when we take the large $N$ limit.

The outline of this paper is as follows: Sections \ref{HiggsD} and
\ref{HiggsE} are devoted to the study of Higgsing RG flows from $D$-
and $E$-type theories, respectively.\footnote{ Higgsing of the
$\widehat{A}$ or $A_{k+1}$ theory only leads to a product theory
without interactions between different factors. This will be
explained in appendix \ref{HiggsA}.} We will mainly consider the
breaking pattern $SU(N)\to SU(N_1)\times SU(N_2)$. The authors of
\cite{KPS} studied the Higgsing RG flows of the $A$-type conformal
theories, where the bifundamental fields are necessarily massive and
get integrated out, leaving two copies of theories without
interactions between them. Here we will consider Higgsing the
theories with the $D_{k+2}$ and $E_7$ superpotentials. We will see
that by tuning the vacuum expectation values, it is possible to
maintain an interaction between the two sectors associated with
$SU(N_1)$ and $SU(N_2)$. Each of $D$ and $E$ theories also admits
several Higgsing RG flows depending on the choice of vacuum
expectation values. In section \ref{HiggsD}, we will focus on one
flow of the ${D_{k+2}}$ theory, which was discussed by Brodie
\cite{Brodie}. By utilizing the electric-magnetic duality, we will
explicitly compute the parameter range (i.e., the conformal window)
where the theory is at a non-trivial fixed point. In section
\ref{HiggsE} we will consider Higgsing of the two-adjoint $SU(N)$
gauge theories with the $E_7$ superpotential. We will study the
non-Abelian Coulomb phase of the resulting $SU(N_1)\times SU(N_2)$
theory that has adjoints, fundamentals for each group and
bifundamental fields together with a superpotential constructed from
bifundamentals. In all cases, we will be able to verify the validity
of the $a$-theorem conjecture.

Sections \ref{MesonO}-\ref{MesonD} can be read
independently of sections \ref{HiggsD} and \ref{HiggsE}. In section
\ref{MesonO} we will study deformations of $\widehat{O}$. In
addition to the  quark mass term $\widetilde{Q}_iQ^i$, we have one
mesonic relevant operator $\widetilde{Q}_i X Q^i$. It
drives the theory to a new fixed point that we call $\widehat{M}$.
In this flow  no gauge invariant operator hits the unitarity bound.
Mass deformation drives the theory to a two-adjoint gauge theory
without $Q_i$. It is also an asymptotically free theory, and we
expect that it flows to an interacting IR fixed point. We call this
new fixed point $M^{\widehat{M}}_{(0,0)}$. We also consider further
RG flows driven by terms like $\tr X^mY^n$ and mesonic terms away
from these new fixed points. The list of the new fixed points that
will appear in section \ref{MesonO} is the following\footnote{ The
fixed points are named in the following way up to renameing
 of $X$ and $Y$: $M$ indicates the
inclusion of a mesonic term in the superpotential. The subscripts
specify the types of superpotentials. While the subscripts ${(m,n)}$
stand for $\tQ X^m Y^nQ$ for the $D$ and $E$ cases, for the $A$ case
$m$ and $n$ stand for number of quark-antiquark fields and that of adjoint field $X$. In the third line of
eq.(\ref{mesondef}), the superpotential takes this form after
integrating out $X$. For the $D$ and $E$ cases when we include more
than two quark fields, we specify the fixed points by ${(s,m,n)}$.
$s$ is the total number of $Q$ and $\widetilde{Q}$.}:
\begin{eqnarray}
\widehat{M}\quad \qquad \qquad W\!\!\! &=& \!\!\! \tQ_i XQ^i \nonu \\
{M}_{(0,0)}^{\widehat{M}} \qquad \qquad W\!\!\!&=&\!\!\! \tQ_i Q^i \nonu \\
{M}^{\widehat{A}}_{(4,0)} \qquad  \qquad W \!\!\! &=&\!\!\! \tQ_i XQ^i+\tr X^2 \nonu \\
{M}^{\widehat{D}}_{(0,1)} \qquad \qquad W\!\!\! &=&\!\!\! \tQ_i XQ^i+\tr X^2Y  \label{mesondef}\\
{M}^{\widehat{E}}_{(1,0)} \qquad \qquad W\!\!\! &=&\!\!\! \tQ_i XQ^i+\tr Y^3 \nonu \\
{M}^{\widehat{E}}_{(0,1)} \qquad \qquad W\!\!\! &=&\!\!\! \tQ_i
XQ^i+\tr X^3 \nonu
\end{eqnarray}

In sections \ref{MesonE}, \ref{MesonA}, and \ref{MesonD}, we will
consider deformations of the $\widehat{E},\widehat{A}$, and
$\widehat{D}$ theories, respectively. We begin with $\widehat{E}$ in
section \ref{MesonE} since it has only a finite number of
interacting RG fixed points and is relatively easy to treat.
The new interacting fixed points to be found in section
\ref{MesonE} are the following:\
\begin{eqnarray}
{M}^{\widehat{E}}_{(k,0)}\qquad  \qquad W \!\!\! &=&\!\!\! \tQ_i X^kQ^i+\tr Y^3,\qquad k=0,1,2,3 \nonu \\
{M}^{\widehat{E}}_{(l,1)}\qquad  \qquad W \!\!\! &=&\!\!\! \tQ_i X^lYQ^i+\tr Y^3,\qquad l=0,1 \label{MhatEsymbols} \\
{M}^{\widehat{E}}_{(4,0,0)}\!\!\!\! \qquad \qquad W\!\!\! &=&\!\!\!
\tQ_i Q^i \tQ_j Q^j+\tr Y^3 \nonu
\end{eqnarray}
In section \ref{MesonA} we will consider the mesonic term
deformations of $\widehat{A}$ and obtain an infinite series of
manifold of fixed points. The RG flows to these fixed points are all
driven by mesonic operators that include more than two quark
superfields:
\begin{eqnarray}
{M}^{\widehat{A}}_{(4,k)}\qquad  W \!\!\! &=&\!\!\! (\tQ_i X^a Q^i) \tQ_j X^bQ^j +(\tQ_j X^a Q^i) \tQ_i X^bQ^j +\tr
Y^2,
\end{eqnarray}
where $a+b=k$, and $k=0,1,2\cdots $. The different contractions of
flavor indices give rise to manifolds of fixed points. In section
\ref{MesonD} we will move on to meson deformations of $\widehat{D}$
and $D_{k+2}$. The $D$-series provide the largest class of CFT's
though the basic features are the same as in the $A$-series. There
are four types of fixed points:
\begin{eqnarray}
{M}^{\widehat{D}}_{(0,k,2)}\qquad  \qquad W \!\!\! &=&\!\!\! \sum(\tr X^a Y) \tQ_jX^bYQ^j+\tr X Y^2, \nonu \\
{M}^{\widehat{D}}_{(0,k,1)} \qquad \qquad W\!\!\! &=&\!\!\! \sum(\tr X^a)  \tQ_jX^bY Q^j+\tr XY^2 .\\
{M}^{\widehat{D}}_{(1,k,0)}\qquad  \qquad W \!\!\! &=&\!\!\! \sum(\tQ_i X^a Q^i ) \tQ_jX^bQ^j+\tr X Y^2, \nonu \\
{M}^{\widehat{D}}_{(1,k,1)} \qquad \qquad W\!\!\! &=&\!\!\! \sum(\tQ_i X^a
Q^i ) \tQ_jX^bY Q^j+\tr XY^2, \nonu
\end{eqnarray}
where the sums are over $a$ and $b$ with $a+b=k$.

In appendix \ref{HiggsA}, we will see why Higgsing the A-type
theories does not lead to new interacting fixed points.

In appendix \ref{exception} we will consider one subtle case in
which it is  not possible to determine which of the two possible
flows actually occurs by a deformation.

%In appendix \ref{curve} we will study the one-adjoint SQCD with a
%mesonic superpotential in terms of a hyperelliptic curve proposed by
%Kapustin \cite{Kapustin}. We will comment on a connection between an
%Argyres-Douglas point obtained from the curve and the non-Abelian
%Coulomb phase. We will be able to reproduce the exact $R$ charges in
%the $A_{k+1}$ conformal theories from the point of view of a
%singular curve.

%%%%%%%%%%%%%%%%%%%%%%%%%%%%%%%%%%%%%%%%%%%%%%%%%%%%%%%%%%%%%%%%%%
\vspace{0.3cm}
\noindent
{\bf Generalities of $\Ncal=1$ RG flows}
\vspace{0.2cm}
%%%%%%%%%%%%%%%%%%%%%%%%%%%%%%%%%%%%%%%%%%%%%%%%%%%%%%%%%%%%%%%%%%

We make use of a version of non-renormalization theorem:
In the presence of sufficient massless matter\footnote{
For the precise statement,
see p.16 of \cite{BIWW}.},
%, so that
%$\sum_i T(r_i)\geq C_2(G)$,
the form of the superpotential is maintained
throughout the RG flow.
The only renormalization comes from renormalization
of the D-terms, including the wave function renormalization.
%({\bf This statement is given in the footnote
%in page 16 of hep-th/0408156. I want to know
%where the proof is given.
%})
In our examples where we have one or two adjoints,
the inequality is always satisfied.

Also, as in the discussion of two-dimensional
$\Ncal=2$ Landau-Ginzburg models,
we assume that a fixed point is characterized by
the superpotential and that the D-terms are automatically
adjusted at a fixed point.

The equations of motion for matter fields
imply that partial derivatives of the superpotential
can be written as $\bar{D}^2(...)$.
The chiral ring is the ring of operators where
partial derivatives of the superpotential are
considered trivial.
Any term in the superpotential that is trivial in the chiral ring
can be then written as a D-term.
The assumption of the previous paragraph implies that
we can use the chiral ring relations to classify relevant
operators.
\footnote{There is an exception
to the assumption.
The deformation by a term proportional to $\p_i W$ is in general relevant,
though $\p_i W$ is trivial in the chiral ring.
This is because the deformation by $\p_i W$ is equivalent to
giving a vev to the chiral field $\Phi_i$, moving to a different
super-selection sector.
Flows $A_k\ra A_{k-1}$ of LG models are of this type.
We will include this type of deformations in our classification.
\label{exceptionfoot}
}

Classification of relevant operators can be further simplified
by considering field redefinitions.

In the case the superpotential has only cubic terms,
the existence of fixed points we find can be confirmed
for large values of $N/N_f$ (or $N_i/N_f$) by perturbation
theory.
We assume that for lower values of $N/N_f$,
these fixed points remain to exist.

In sections 4-7 we will study  deformations
by a product of gauge invariant chiral operators
${\cal O}_1, {\cal O}_2, \cdots$.
The unitarity bound requires that each chiral primary has dimension no more than one.
Since a chiral operator is relevant if and only if it has dimension less than three,
we see that a relevant chiral operator is a product of
at most two gauge invariant operators.

%%%%%%%%%%%%%%%%%%%%%%%%%%%%%%%%%%%%%%%%%%%%%%%%%%%%%%%%%%%%%%%%%%%%
\section{Higgsing of $D$-type Fixed Points}\label{HiggsD}
%%%%%%%%%%%%%%%%%%%%%%%%%%%%%%%%%%%%%%%%%%%%%%%%%%%%%%%%%%%%%%%%%%%%

In this section, we consider Higgsing the $D$-type theories.
Allowed patterns of gauge symmetry breaking
are determined by $D$- and $F$-term
conditions. For simplicity we focus on the
breaking pattern $U(N)\to U(N_1)\times U(N_2)$\footnote{ We
work with $U(N)$ rather than $SU(N)$ in sections \ref{HiggsD} and
\ref{HiggsE} as done in
\cite{IWADE}. The overall $U(1)$ gauge field decouples in the IR
because the beta function is positive. In the absence of a
superpotential, the trace parts of the adjoints are free, and in the
large $N$ limit can be ignored when computing the central charge
$a$. Even in the presence of a superpotential that couples the trace
parts to the rest of the theory, we expect that there is not much
difference between the $U(N)$ and $SU(N)$ cases. } though more complicated patterns can
be analyzed similarly. Since we are interested in a region
where both gauge groups are asymptotically free, $x\equiv N_1/N$ and
$y\equiv N_2/N$ must satisfy the inequalities $2x-1\ge y\ge
{x+1\over 2}$. The theory is invariant under $N_1\leftrightarrow
N_2$ and we restrict ourselves to the region $y\le x$.

As discussed in \cite{Brodie}, there are two kinds of Higgsing of
the $D_{k+2}$ theory. In one case we have $\langle X \rangle = 0$
and $\langle Y \rangle \ne 0$, where the theory is driven in the IR
to the CFT studied by Intriligator, Leigh, and Strassler in
\cite{ILS}. This CFT has a superpotential $\tr
(F\widetilde{F})^{k+1}$ constructed from bifundamentals $F$ and
$\widetilde{F}$, and does not contain adjoint fields. The
non-Abelian Coulomb phase of this model was studied in detail in
\cite{Intriligator0504} with the help of $a$-maximization. They showed
that the RG running of the multiple couplings can affect each other
and found several interacting RG-fixed points and non-empty
superconformal windows by using the electric and magnetic
descriptions. We do not touch upon this model any
further.

On the other hand the expectation values $\langle Y \rangle
=0$ and $\langle X \rangle \ne 0$ drive the theory to the model
studied by Brodie, who also proposed a dual description
\cite{Brodie}.
We are  interested in finding the conformal window of his model,
and for this purpose it is useful to start with Higgsing of
$\widehat{D}$ theory and then to look for a region where $\tr
X_{1}^{k+1}$ and $\tr X_2^{k+1}$ are relevant. Thus we will show
breaking pattern of $\widehat{D}$ theory.
In order
to break the original gauge group $U(N)$ into two parts we consider
the vev
\begin{eqnarray}
\langle X \rangle ={\rm
diag}(\stackrel{N_1}{\overbrace{a,\cdots,a}},
\stackrel{N_2}{\overbrace{b,\cdots,b}}),\qquad N_1+N_2=N.
\label{VEVX}
\end{eqnarray}
Bifundamental fields coming from the fluctuations of $X$ become
massive through $D$-terms and are integrated out although the
adjoints $X_i$ of $U(N_i)$, coming from the block diagonal parts of
X, remain massless. Since $\langle Y \rangle=0$, the $D$-terms do
not give mass to the bifundamentals $F$ and $\widetilde{F}$ coming
from the fluctuations of $Y$. The superpotential $\tr XY^2$,
however, gives rise to the terms $\tr a {\bf
1}_{N_1}F\widetilde{F}+\tr b{\bf 1}_{N_2}\widetilde{F}F=(a+b)\tr
F\widetilde{F}$ that can make the bifundamentals massive.
We let $b=-a$ to keep $F$ and $\tilde{F}$ massless,
maintaining an interaction between the two gauge groups in the IR.
We finally obtain a $U(N_1)\times U(N_2)$ gauge theory with
massless fields $X_1,X_2$, $F,\widetilde{F}$ and $N_f$ quark,
anti-quark superfields $(\widetilde{Q}_1,Q_1)$ and
$(\widetilde{Q}_2,Q_2)$ for both gauge groups\footnote{We suppress
 the flavor indices of $Q_{1,2}$ and $\widetilde{Q}_{1,2}$,
 in addition to the gauge indices that have been also omitted.},  together
with the superpotential
\begin{eqnarray}
W=\tr X_1F\widetilde{F}+ \tr X_2\widetilde{F}F.  \label{elecpot}
\end{eqnarray}
We now study the non-Abelian Coulomb phase of this theory and then
deform it by $\tr X_1^{k+1}+\tr X_2^{k+1}$.
Brodie's theory has a
dual description, and we propose that the theory is conformal only in
the parameter region where, when we consider a theory without the
superpotentials, the superpotentials we just removed are relevant
deformations.
This criterion is allows us to determine the conformal window of
Brodie's theory. In particular we will explicitly show the conformal
window for the model with $k=4$ and large $k$. Also we will check
the $a$-theorem under the Higgsing RG flows.

%%%%%%%%%%%%%%%%%%%%%%%%%%%%%%%%%%%%%%%%%%%%%%%%%%%%%%%%%%%%%%%%%%%
\subsection{Electric description}
%%%%%%%%%%%%%%%%%%%%%%%%%%%%%%%%%%%%%%%%%%%%%%%%%%%%%%%%%%%%%%%%%%%

First we study the electric description, which has the superpotential
(\ref{elecpot}). Taking into account the marginality of the
superpotential terms $R(X_1)=R(X_2)=2-2R(F)$ and the vanishing of
the ABJ anomaly
\begin{eqnarray}
(R(Q_1)-1)+xR(X_1)+y(R(F)-1)=(R(Q_2)-1)+yR(X_2)+x(R(F)-1)=0, \label{Rchargerelations}
\end{eqnarray}
we are left with one undetermined variable corresponding
one flavor $U(1)$ symmetry that mixes with
the $R$ symmetry in the IR.

\begin{figure}[htbp]
\begin{center}
\includegraphics[width=15.0cm,height=3.8cm]
{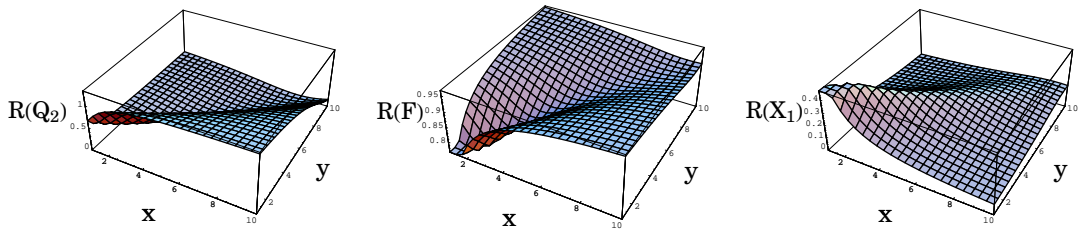}
\end{center}
 \caption{\small  $R$ charges of $Q_2$, $F$ and $X_1$ without taking into account the operators
 hitting the unitarity bound.}
    \label{RchargesQ2FX1}
\end{figure}

The trial $a$-function, in the convention eq.(\ref{aconvention}), is
\begin{eqnarray}
a^{(0)}=\!\!\!\!&&2x^2+2y^2+x^2[3(R(X_1)-1)^3-(R(X_1)-1)]+y^2[3(R(X_2)-1)^3-(R(X_2)-1)]\nonu \\
&&+2xy[3(R(F)-1)^3-(R(F)-1)]+2x[3(R(Q_1)-1)^3-(R(Q_1)-1)]\nonu \\
&&+2y[3(R(Q_2)-1)^3-(R(Q_2)-1)].
\end{eqnarray}
By maximizing this function we determine the $R$ charges of the
fields and the value of the central charge $a$.
$R(X_1)$ is
\begin{eqnarray}
&&R(X_1)= \frac{20 x^2-2 y x+20 y^2}{3 \left(6 x^2+6
   y^2+\sqrt{F[x,y]+F[y,x]}\right)}, \\
&{\rm where}&F[x,y]\equiv \frac{x^2}{2} \left(160 x^4-276 y x^3+426 y^2 x^2-8
   x^2-272 y^3 x+28 y x-9 y^2\right). \nonu
\end{eqnarray}
Other $R$
charges are obtained from this through the relations above
and are displayed in figure \ref{RchargesQ2FX1}.
%%%%%%%%%%%%%%%%%%%%%%%%%%%%%%%%%%%%%%%%%%%%%%%%%%%%%%%
%%%%%%%%%%%%%%%%%%%%%%%%%%%%%%%%%%%%%%%%%%%%%%%%%%%%%%%%

Depending on the values of $x$ and $y$, some operators turn out to
have $R$ charges below the free field value $2/3$. These operators
should have been decoupled at lower values of $x$ and $y$.
We propose a prescription, generalizing the one in
\cite{KPS}, for decoupling these operators.
Consider straight lines
on the $xy$ plane starting from the point $(x,y)=(1,1)$ at various
angles. Moving to the right and up along such a line from the point
$(1,1)$, $\widetilde{Q}_1 Q_1$ hits the unitarity bound first.
Beyond this point along the line, we need to modify the trial
$a$-function by subtracting the contribution of the decoupled
operator and adding the contribution from a free field. As we
proceed further on the line, more and more operators hit the
unitarity bound and get decoupled, and each time this happens, we
modify the trial $a$-function (see eq.(\ref{electa})).

\begin{figure}[htbp]
\begin{center}
\includegraphics[width=6.5cm,height=5.0cm]
{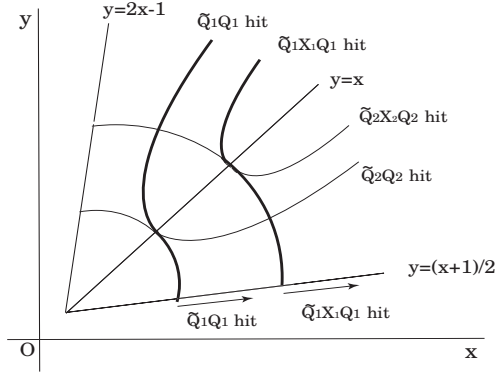}
\end{center}
    \caption{\small Qualitative picture of operators hitting the unitarity bound.
    Only $P_{l,1}=\widetilde{Q}_1 X^{l-1}_1 Q_1$ and $M_{l,1}=\widetilde{Q}_2 X^{l-1}_2 Q_1$ can saturate the bound.}
    \label{RoughFig}
\end{figure}

We can qualitatively see which operators hit the unitarity bound and
get decoupled from the results in figure \ref{RchargesQ2FX1}, by
ignoring the modifications to the trial $a$-function. Our theory has
several types of mesonic operators defined by\footnote{ Our notation
here is different from that in \cite{Brodie}.}
$P_{l,1}=\tQ_1X_1^{l-1}Q_1$, $M_{l,1}=\tQ_2X_2^{l-1}{Q}_2$,
$P_{l,2}=\tQ_1X_1^{l-1}F{Q}_2$,
$M_{l,2}=\tQ_2X_2^{l-1}\widetilde{F}{Q}_1$,
$P_{l,3}=\tQ_1X_1^{l-1}F\widetilde{F}{Q}_1$, and
$M_{l,3}=\tQ_2\widetilde{F}FX_2^{l-1}{Q}_2$. Using the results above
we see that the $R$ charges of the bifundamentals are greater than
${2\over 3}$ everywhere.
The $R$ charges of the other chiral fields are positive,
except that the $R$ charges of $Q_1$ and $Q_2$ 
become only slightly negative at large values of $x$ and $y$.
Thus the mesonic operators that contain
bifundamentals do not cross the unitarity bound, and only $P_{l,1}$
and $M_{l,1}$ can reach the bound. The region where the operators
hit the bound is qualitatively depicted in figure \ref{RoughFig}.

%%%%%%%%%%%%%%%%%%%%%%%%%%%%%%%%%%%%%%%%%%%%%%%%%%%%%%%
%%%%%%%%%%%%%%%%%%%%%%%%%%%%%%%%%%%%%%%%%%%%%%%%%%%%%%%%

We now follow the precise procedure described above and compute the
$R$ charges exactly, including the back-reaction of the decoupled
operators. 
We subtract from the trial $a$-function the contributions
of the first $m$ $P_{l,1}$ and $n$ $M_{l,1}$ that hit the bound,
and then add the corresponding contributions from as many free
fields with the relation:
\begin{eqnarray}
a^{(m,n)}\equiv a^{(0)}+{1\over
9}\sum_{l=1}^{m}[2-3R(P_{l,1})]^2[5-3R(P_{l,1})]+ {1\over
9}\sum_{l=1}^{n}[2-3R(M_{l,1})]^2[5-3R(M_{l,1})] ,\label{electa}
\end{eqnarray}
where the $R$ charge of each operator can be expressed in terms of
$R(Q_1)$,
\begin{eqnarray}
&&R(P_{l,1})=2R(Q_1)+(l-1){2-2R(Q_1)\over 2x-y},\nonu \\
&&R(M_{l,1})=2\left({x-2y\over y-2x}(R(Q_1)-1)+1
\right)+(l-1){2-2R(Q_1)\over 2x-y}.
\end{eqnarray}
We also used the fact that
\begin{eqnarray}
-[3(R-1)^3-(R-1)]+[3(\frac{2}{3}-1)^3-(\frac{2}{3}-1)]=\frac{1}{9}(2-3R)^2(5-3R) \nonu
\end{eqnarray}
in writing eq.(\ref{electa}).

%Though in principle possible, it is difficult to study all the allowed region on the $xy$-plane.
%We focus on the asymptotic behaviors of the $R$ charges and the  central charge $a$
%on the lines $y=x$ and $y={x+1\over 2}$.
%From the results
%we will see where the operator $\tr X_1^{k+1}+\tr X_2^{k+1}$ becomes
%relevant at large $x$, which give a ``an edge of the window''.
$a^{(0)}$ can be evaluated at large values of $x$ and $y$:
\begin{eqnarray}
a^{(0)}\simeq \!\!\!\!\!\!&&{1\over (2 x-y)^3}\left[2 ( z-1) \left(8
\left(3 z^2-6 z-2\right) x^4+\left(-39 y
    z^2+78 y z+5 y\right) x^3 \right. \right. \nonu \\
&&\left. \left. +18 y \left(2 y z^2-4 y
   z+z-y\right) x^2+y^2 \left(-39 y z^2+78 y z+2
   y\right) x \right. \right. \nonu \\
&&\left. \left. +2 y^3 \left(12 y z^2-24 y z+7
y\right)\right)\right] ,
\end{eqnarray}
where $z\equiv R(Q_1)$.

As we see from figure \ref{RoughFig}, a large number of $P_{l,1}$
hit the unitarity bound at large values of $x$ and $y$. In this
limit, as in \cite{KPS}, we can approximate the first sum in eq.(\ref{electa}) by an integral:
%In the large $N$ limit we can replace the sum over $l$ with an integral
%because $R(P_{l,1})$ changes continuously in the large $x,y$ limit and then obtain,
\begin{eqnarray}
{1\over 9}\sum_{l=1}^{m}[2-3R(P_{l,1})]^2[5-3R(P_{l,1})]\simeq
{1\over 27\beta}\int_{0}^{2-3\alpha}u^2(3+u)du ={2\over
9}(2x-y)(1-3R(Q_1))^3,
\end{eqnarray}
where $u\equiv 2-3R(P_{l,1})$, $\alpha \equiv 2R(Q_1)$ and $\beta
\equiv {2-2R(Q_1)\over 2x-y}$. The situation is a little different
for $M_{l,1}$. As seen from figure \ref{RoughFig}, not for all
ratios of $x$ and $y$ do the operators $M_{l,1}$ hit the unitarity
bound, making it difficult to calculate the second sum in eq.(\ref{electa}) in general. For this reason we will consider two
limiting cases where the number of $M_{l,1}$ hitting the unitarity
bound is either zero or very large.

The asymptotic forms of the $R$ charges and the central charge in
the region near the line $y=(x+1)/2$, where only $P_{l,1}$ hit the
unitarity bound, can be explicitly computed. The expressions right
on the line $y={x+1\over 2}$ are simple to write and are given as
as $R(Q_1)\simeq  -0.057$, $R(X_1)\simeq 1.409x^{-1}$, and $
a^{y=(x+1)/2} \simeq 10.36x$. Using these asymptotic values we see
that $\tr X^k$ becomes relevant in the region $x>x_k=0.704k$. This
provides one end of the conformal window on $y=(x+1)/2$. (The other
is to be found from the analysis of the magnetic dual.)

Finally we will show that baryonic operators do not violate the
unitarity bound on the line $y={x+1\over 2}$. To construct a
baryonic operator we need enough number different quark superfields
including dressed quark $(Q_{(I_1\cdots I_n)})_i\equiv
(X_{I_1}\cdots X_{I_n}Q)_i$. At a point $x=[x]$ we need $N=[x]N_f$
number of quarks to contract with indices of epsilon tensor.
Therefore at least $([x]-1)N_f$ number of $X_i$ are included in
baryonic operators. It is enough to consider the operator $B_{\rm
small}\equiv \prod_{i=0}^{[x]-1}(X_1^iQ_1)^{N_f}$ because this gives
the smallest $R$ charge possessed by the baryonic operators. Since
the order of the $R(B_{\rm small})$ is ${\cal O}(N)$, either
$R(Q_1)$ or $R(X_1)$ have to be zero or negative to hit the bound.
This can happen in the large $x$ region. Representing the
$R(X_1)$ in terms of $R(Q_1)$ we have $R(B_{\rm small})\simeq N (
R(Q_1)+[x](1-R(Q_1)))$. If we have a unitarity violation the
condition, $R(Q_1)+[x](1-R(Q_1))\simeq 0$ have to be satisfied.
Since $R(Q_1)$ varies within $-0.057 \le R(Q_1)\le 1$, however, it
never hold. Therefore we conclude that there is no baryonic operator
that hits the unitarity bound.

Next let us study the asymptotic behavior of the $R$ charges on the
line $y=x$. If we restrict ourselves to the line $y={x}$ and large
values of $x$, a large number of $M_{l,1}$ hit the bound. We obtain
by the same approximation
\begin{eqnarray}
{1\over
9}\sum_{l=1}^{m}[2-3R(M_{l,1})]^2[5-3R(M_{l,1})]\Big|_{y=x}\simeq
{2\over 9} x(1-3R(Q_1))^3. \label{aa20}
\end{eqnarray}
Taking into account eqs. (\ref{electa})-(\ref{aa20}), we obtain the
asymptotic values of the $R$ charges $R(Q_1)\simeq -{1\over 8}$,
$R(X_1)={9\over 4x}$, and the central charge $a^{y=x} \simeq
26.81x$. One might worry again that baryonic operators may  hit the
unitarity bound. Proceeding in the same way as in the previous case,
however, one can see that there is no  baryonic operator hitting the
bound. We conclude that the operator $\tr X_1^{k+1}$ becomes
relevant in the region $x>x_k\equiv {9k\over 8}$.

Now that we have obtained the asymptotic behavior of the central
charge $a$, we would like to test the $a$-theorem conjecture under
the RG flow from $\widehat{D}$. In so doing we have to be careful
about the arguments of functions. In \cite{IWADE}, the central
charge $a$ of the $\widehat{D}$ theory was expressed in terms of the
variable $\widehat{x}\equiv {N\over N_f}$ while we are using
$x={N_1\over N_f}$ and $y={N_2 \over N_f}$ that satisfy the relation
$\widehat{x}=x+y$. Thus comparing these results we have to use the
variable $\widehat{x}$. Taking this point into account,  we obtain
\begin{eqnarray}
a^{y=(x+1)/2}(x)\Big|_{x={2\over 3}(\widehat{x}-{1\over 2})}\simeq
6.906 \widehat{x}, \qquad
a^{y=x}(x)\Big|_{x=\widehat{x}/2}\simeq 13.40\widehat{x}.
\end{eqnarray}
The central charge of the  $\widehat{D}$ theory was given in
\cite{IWADE} as $a_{\widehat{D}}(\widehat{x}) \simeq
13.40\widehat{x}$. By comparing these central charges we see that
the $a$-theorem is satisfied under the RG flow:
\begin{eqnarray}
a_{\widehat{D}}=a^{y=x}>a^{y=(x+1)/2}.
\end{eqnarray}

We see that the central charge on the line $y=x$ is equal to that of
$\widehat{D}$ when $x$ is large. This is explained as follows. Let
us assume that $N_1=N_2$, without assuming that $x=y$ is large. By
comparing the superpotential eq.(\ref{elecpot}) with $\tr XY^2$ of
$\widehat{D}$, we see that $X_1$ and $X_2$ play the role of $X$ in
$\widehat{D}$, while $F$ and $\widetilde{F}$ play the role of $Y$,
as far as the calculation of the trial $a$-function is concerned.
Therefore central charge $a_{Higgsing}$ of the theory after Higgsing
can be written as
$a_{Higgsing}(N_1,N_2=N_1)=2a_{\widehat{D}}(N=N_1)$ even before
$a$-maximization. This relation translates to the relation
$a_{Higgsing}(x,y=x)=2a_{\widehat{D}}(\widehat{x}=x)$, which holds
even for finite values of $x$. Since in \cite{IWADE} the central
charge $a_{\widehat{D}}$ of the $\widehat{D}$ theory was computed
for finite values $x$, we can translate the result there to the
central charge $a_{Higgsing}(x,y=x)$ of the theory after Higgsing
with $x=y$. In the above we saw that $a_{Higgsing}(x,y=x)$ is linear
in $x$ when $x$ is large. In this limit, we then have
$a_{Higgsing}(x,y=x)\simeq 2^{-1}a_{Higgsing}(2x,y=2x)
=a_{\widehat{D}}(\widehat{x}=2x)$ as we saw above.

\begin{figure}[htbp]
\begin{center}
\includegraphics[width=4.5cm,height=2.5cm]
{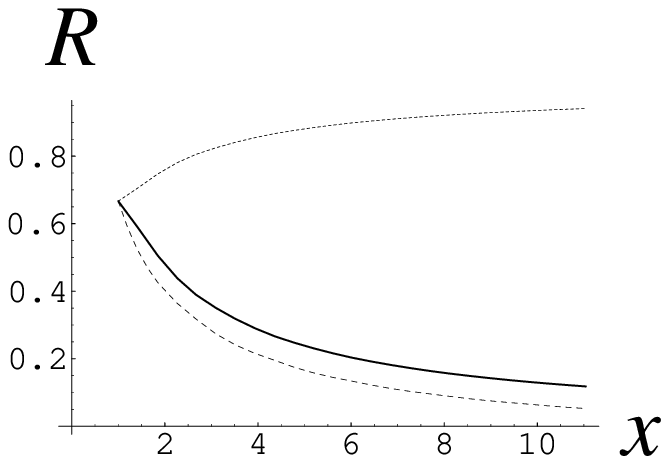}\hspace{0.7cm}
\includegraphics[width=4.5cm,height=2.5cm]
{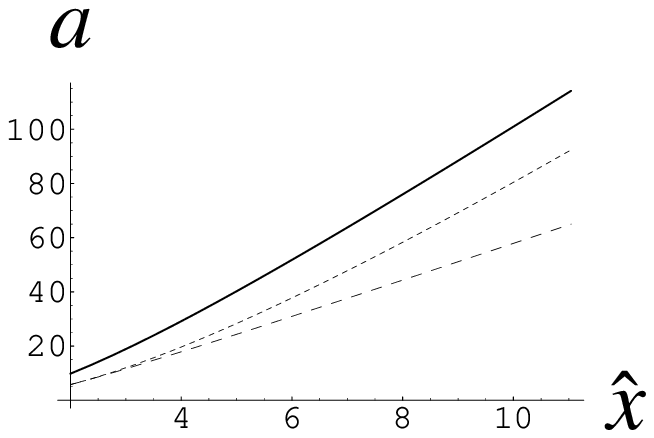}
\end{center}
    \caption{\small The figure on the left displays the $R$ charges $R(Q_1)$ (top), $R(Q_2)$ (middle),
    and $R(X_1)$ (bottom) on the line $y={x+1 \over 2}$. The figure on the right  depicts
    the central charge of the $\widehat{D}$ theory (top) as well as  those of the theories after Higgsing
    on the lines $y=x$ (middle) and $y={x+1\over 2}$ (bottom).  }
    \label{hitcase}
\end{figure}

%As we saw in the previous paragraph, we can learn the value of
%the central charge  after Higgsing, even for finite values of $x$,
%by carrying over the results for $\widehat{D}$ in
%\cite{IWADE}, provided that $y=x$.
As for the line $y={x+1\over 2}$ we have to calculate
the central charge in detail, by following the procedure proposed
early in this subsection. On this line only $P_{l,1}$ can hit the
unitarity bound. For $x\gtrsim 1$,
we take $a^{(0)}$ and maximize it. The obtained $R$ charges are
correct up to $x\simeq 2.54$, where $P_{l=1,1}$ hits the unitarity
bound. Above this value of $x$ we switch to $a^{(1,0)}$ and maximize it. Repeating the
process in this way we see, for example, that the operators
$P_{l,1}$ from $l=2$ to $6$ hit the bound  at $x\simeq \ 4.29,\
6.08,\ 7.88,\ 9.68$ and $ 13.38$. Patching the results together, we
obtain the $R$ charges and the central charge up to ${N/N_f}=
13.38$. The central charge of the $\widehat{D}$ theory, as well as
those of the theories after Higgsing with $y=x$ and $y=(x+1)/2$ are
depicted in figure \ref{hitcase}. Clearly the results indicate the
validity of the $a$-theorem conjecture in these Higgsing RG flows.

%%%%%%%%%%%%%%%%%%%%%%%%%%%%%%%%%%%%%%%%%%%%%%%%%%%%%%%
%%%%%%%%%%%%%%%%%%%%%%%%%%%%%%%%%%%%%%%%%%%%%%%%%%%%%%%%

For later reference we calculate the value of $x$ at which
$\tr X^4_1$ and $\tr X^4_2$ become relevant on the line $y={x+1\over 2}$.
It turns out to be $x=1.879$,
where no operator has decoupled by hitting the unitarity bound.

%%%%%%%%%%%%%%%%%%%%%%%%%%%%%%%%%%%%%%%%%%%%%%%%%%%%%%%%%%%%%%%%%%%%%%%%%%%%%
\subsection{Analysis of the magnetic dual description}
%%%%%%%%%%%%%%%%%%%%%%%%%%%%%%%%%%%%%%%%%%%%%%%%%%%%%%%%%%%%%%%%%%%%%%%%%%%%%

\begin{figure}[htbp]
\begin{center}
\includegraphics[width=11.5cm,height=4.0cm]
{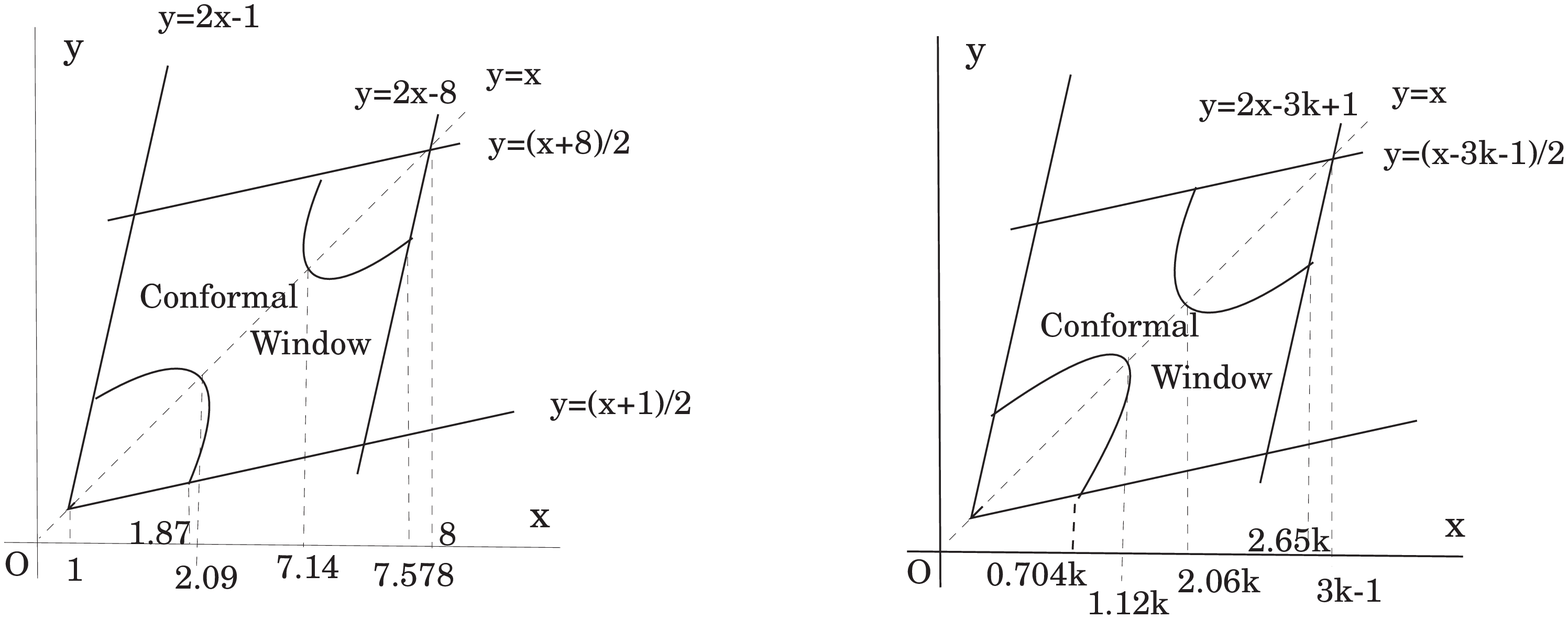}
\end{center}
    \caption{\small
    Conformal windows of the theories with $k=3$ and $k\gg 1$.
    In each graph, we computed the numerical values of $x$ at the corners of the
    conformal window and interpolated between the corners by hand. }
    \label{CWindow}
\end{figure}

The magnetic dual description of the present model was proposed by
Brodie in \cite{Brodie}. The dual gauge group is $U(3kN_f-N_1)\times
U(3kN_f-N_2)$. The matter contents include $\bar{X}_1,\bar{X}_2,
\bar{F}$, and $\widetilde{\bar{F}}$ that correspond to $X_1,X_2,F$,
and $\widetilde{F}$ in the electric description, respectively. In
addition, there are dual quarks $q_1$ and $q_2$, and the gauge
singlets $P_{l,m}$ and $M_{l,m}$ ($m=1,2,3$) to be identified with
the mesonic operators which have appeared in the electric
description. The superpotential is \cite{Brodie}
\begin{eqnarray}
W=&&\!\!\!\!\!\! \tr \bar{X}_1^{k+1}+\tr \bar{X}_2^{k+1}
+\tr \bar{X}_1\widetilde{\bar{F}}{\bar{F}}+\tr \bar{X}_2\widetilde{\bar{F}}\bar{F}+\sum_{l=1}^{k}\left[ M_{l,1}\widetilde{q}_2\bar{X}_2^{k-l}\bar{F}\widetilde{\bar{F}}q_2+P_{l,1}\widetilde{q}_1\bar{X}_1^{k-l}\widetilde{\bar{F}}\bar{F}q_1 \right. \nonu \\
&&\left.+P_{l,2}\widetilde{q}_2\bar{X}_2^{k-l}\bar{F}q_1
+M_{l,2}\widetilde{q}_1\bar{X}_1^{k-l}\widetilde{\bar{F}}q_2+M_{l,3}\widetilde{q}_2\bar{X}_{2}^{k-l}q_2+P_{l,3}\widetilde{q}_1\bar{X}_1^{k-l}q_1
\right]. \label{AllDMagW}
\end{eqnarray}
We define a new notation
$\overline{N}_1\equiv 3kN_f-N_1$, $\overline{N}_2\equiv 3kN_f-N_2$,
$\bar{x}\equiv {\overline{N}_1\over N_f}$ and $\bar{y}\equiv
{\overline{N}_2\over N_f}$. Since we are interested in the parameter
region where both factors of the gauge group in the magnetic theory
are asymptotically free, we consider the following region;
\begin{eqnarray}
2\bar{x}-1\ge \bar{y}\ge {(\bar{x}+1)\over 2} \iff  {x+3k-1\over 2}
\ge y \ge 2x-3k+1.
\end{eqnarray}
However since it is difficult to study the whole region, we will
mostly focus on the lines $y=x$ and $y=2x-3k+1$\footnote{
$\bar{X}_1$ remains to have a nontrivial $R$ charge even on the line
$y=2x-3k+1$ where $SU(\overline{N}_1)$ ceases to be asymptotically
free due to the superpotential interaction $\bar{X}_1
\bar{F}\widetilde{\bar{F}}+\bar{X}_2 \bar{F}\widetilde{\bar{F}}$.}.
(See figure \ref{CWindow}.)

Let us begin by studying the line $y=2x-3k+1$. We start from the
point $\bar{x}=\bar{y}=1$ where both factors of the magnetic theory
cease to be asymptotically free and become free in the IR. Thus all
matter content have $R$ charge ${2\over 3}$ \footnote{One might
worry about an effect of a superpotential on the boundary, which
gives rise to $R\ne {2\over3 }$. However one-loop beta function
becomes zero on the boundary. The assumption that NSVZ beta function is
zero forces all the matter fields to have anomalous dimension
$\gamma =0$, or equivalently $R={2\over 3}$. } so only cubic terms in
the superpotential can be marginal near the point,
\begin{eqnarray}
W=\bar{X}_1\bar{F}\widetilde{\bar{F}}+\bar{X}_2\bar{F}\widetilde{\bar{F}}+M_{k,3}\widetilde{q}_2q_2+P_{k,3}\widetilde{q}_1q_1.
\label{DMagW}
\end{eqnarray}
Taking into account the vanishing of the anomaly, we evaluate the
$R$ charges away from the point, $(\bar{x},\bar{y})=(1,1)$. Trial
$a$-function can be written as
\begin{eqnarray}
a&=&2\bar{x}^2+2\bar{y}^2+\bar{x}^2 G[R(X_1)]+2\bar{x}G[R(q_1)]+2\bar{x}\bar{y}G[R(F)] +2\bar{y}G[R(q_2)] \nonu \\
&&+\bar{y}^2G[R(X_2)]+G[R(M_{k,3})]+G[R(P_{k,3})]+{2\over 9}(6k-2)
\label{BrodieTM}
\end{eqnarray}
where we defined $G[x]\equiv 3(x-1)^3-(x-1)$. The last term
comes from the free singlets. Among the all singlets only two are
interacting, thus we have $6k-2$ free singlets. Away from the origin
the $R$ charges of operators vary and some of the terms in the
superpotential become relevant. As for the $k=3$ case, by using
$a$-maximization, we can see  that the operator $\tr \bar{X}_1^4$ and
$\tr \bar{X}_2^4$ becomes marginal at $\bar{x}=1.422$. On the other
hand, looking at $R$ charges of the fields we see that $
M_{k-1,3}\widetilde{q}_2\bar{X}_2q_2$ becomes marginal next at
$\bar{x}=1.452$. Thus before this term becomes relevant $\tr
\bar{X}_1^{4}$ and $\tr \bar{X}_2^{4}$ become relevant and thus the
conformal window starts from $\bar{x}=1.422$ . Combining the results
we can roughly draw the conformal window as in figure \ref{CWindow}.

Next let us study the asymptotic behaviors on this line,
$y=2x-3k+1$. At $(\bar{x},\bar{y})=(1,1)$ the $R$ charges of all
fields are ${2\over 3}$. First since only cubic superpotential can
be marginal, we maximize $a$-function while imposing the marginality
of the superpotential (\ref{DMagW}) and the ABJ anomaly
cancellation. To see the behavior we calculated several values of
$k$ and found that $R(q_1)$ and $R(F)$ are monotonically increasing
functions and take values greater than ${2\over 3}$. On the other
hand $R(q_2)$ and $R(\bar{X}_1)=R(\bar{X}_2)$ are monotonically
decreasing functions. Also all the $R$ charges are positive. Thus
among the superpotential terms (\ref{AllDMagW}) including more than
two of $q_1$ and $F$ can not be relevant. Only
$M_{l,3}\widetilde{q}_2\bar{X}_2^{k-1}q_2$ become relevant and the
point where it becomes marginal is given by the  solution to the
equation,
\begin{eqnarray}
2R(q_2)+(p-1)R(\bar{X}_2)={4\over 3}. \label{conma}
\end{eqnarray}
where $p$ is the number of superpotential terms which are marginal.
In other words we assume that first $k-p$ singlets $M_{1,3},\cdots
M_{k-p,3}$ are free.

First let us calculate asymptotic behavior of the contribution
coming from interacting $M_{l,3}$. Using the marginality of
superpotential we can rewrite the summation of $l=k-p+1,\cdots  k$
by $j=1,\cdots p$ as follows:
\begin{eqnarray}
\sum_{l=k-p+1}^kG[R(M_{l,3})]&=&\sum_{j=1}^pG[2-R(\widetilde{q}_2
\bar{X}_2^{k-l}q_2)] \nonu \\
&=&{1\over 9}\sum_{l=1}^{p}\left(2-3R(N_{l,3})
\right)^2\left(5-3R(N_{l,3}) \right)- {2\over 9}p \label{a3b}
\end{eqnarray}
where we defined $N_{j,3}\equiv \widetilde{q}_2\bar{X}_2^{p-j}q_2$
and the relation noted just above $(2.13)$. From the vanishing of
the anomaly and the marginality of the superpotential
$\bar{X}_2\bar{F}\widetilde{\bar{F}}$, the $R$ charge of $\bar{X}_2$
is written in terms of $R(q_2)$ as $R(\bar{X}_2)={2-2R(q_2)\over
3\bar{x}-2}$. At large $\bar{x}$ variable $u$ defined by
\begin{eqnarray}
u\equiv 2-3R(N_{j,3}) = 2-3\left( 2R(q_2)+(p-j) {2-2R(q_2) \over
3\bar{x}-2}\right)
\end{eqnarray}
becomes continuous. Therefore we can approximate the contributions
of $M_{l,3}$ by an integral. Using (\ref{conma}) we see that $u$
vary within $[-2,2-6R(q_2)]$. Thus contribution of the interacting
$M_{l,3}$ (\ref{a3b}) can be written by an integral as follows:
\begin{eqnarray}
\sum_{l=k-p+1}^kG[R(M_{l,3})]&\simeq & {\bar{x}\over 18(1-R(q_2))} \int_{-2}^{2-6R(q_2)} du\, u^2 (3+u)-{2\over 9}p \nonu \\
&=&{2\bar{x}\over 3}(1-3R(q_2))^3+{2\bar{x}\over
9(1-R(q_2))}-{2\over 9}p. \label{asy1}
\end{eqnarray}
Note that using (\ref{a3b}) we can represent the $p$ in terms of
$R(q_2)$ as $p\simeq {2-3R(q_2)\over 1-R(q_2)}\bar{x}$. Another
contributions to the trial $a$-function can be written as
\begin{eqnarray}
a&=&2\bar{x}^2+2\bar{y}^2+\bar{x}^2 G[R(X_1)]+2\bar{x}G[R(q_1)]+2\bar{x}\bar{y}G[R(F)] +2\bar{y}G[R(q_2)] \nonu \\
&&+\bar{y}^2G[R(X_2)]+G[R(P_{k,3})]+{2\over 9}(6k-1-p) \nonu \\
&\simeq & 12 R(q_2)^3-36R(q_2)^2+4R(q_2)+20+{2\over 9}(6k-p)
\label{asy2}
\end{eqnarray}
Adding (\ref{asy1}) and (\ref{asy2}) we obtain the trial
$a$-function at large values of $\bar{x}$. By maximizing it we
obtain $R(q_2)\simeq -0.0391$. Using this result we see that $\tr
X_1^{k}$ becomes relevant at $\bar{x}\simeq 0.346 k$ equivalently
$x\simeq 2.65k$. Plugging these $R$ charges back into the central
charge gives $a^{mag}\simeq 20.02\bar{x}+{4k\over 3}$.

Let us now turn to the magnetic description on the line $y=x$. By
the same argument as in the previous subsection, the central charge
on this line is related to that of $D_{k+2}$ by
$a^{mag}_{Higgsing}(\overline{N}_1,\overline{N}_2=\overline{N}_1)=2a_{{D}_{k+2}}^{mag}
(\overline{N}=\overline{N}_1)$ if we identify $\overline{N}=3kN_f-N$
in \cite{IWADE} with $\overline{N}_1$ in our model. To rewrite the
central charge $a^{mag}_{Higgsing}$ in terms of
$\widehat{\bar{x}}\equiv {\overline{N}\over N_f}$ we use the
relation $\widehat{\bar{x}}=2\bar{x}-3k$. Therefore by using the
results for the dual of $D_{k+2}$, we obtain the central charge of
our magnetic dual as
$a^{mag}_{Higgsing}(\bar{x}=(\widehat{\bar{x}}+3k)/2,\bar{y}=\bar{x})=
2a_{D_{k+2}}^{mag}(\widehat{\bar{x}})$. To begin with let us
consider the large $\bar{x}$ behavior of the central charge and look
for a point where the operators $\tr \bar{X}_1^{k+1}$ and $\tr
\bar{X}_2^{k+1}$ become marginal. Using the result
$a^{mag}_{D_{k+2}}(\widehat{\bar{x}})\simeq
13.1186\widehat{\bar{x}}+{6k\over 9}$ given in \cite{IWADE}, we
obtain
$a_{Higgsing}^{mag}(\bar{x}=(\widehat{\bar{x}}+3k)/2,\bar{y}=\bar{x})\simeq
13.1186\widehat{\bar{x}}+40.68 k$. The point where $\tr
\bar{X}_1^{k+1}$ becomes relevant is ${\overline{N}_1\over
N_f}\simeq 1.1038k$. To compute the conformal window of the $D_5$
theory, let us see where $\tr X^4_1$ becomes marginal. From
\cite{IWADE} $\bar{x}_5^{min}\simeq 2.09$ thus the conformal window
starts at $3k-1-\bar{x}_{5}^{min}\simeq 7.14$.

Combining the analysis of the electric and magnetic descriptions, we
obtain the conformal window in the large $k$ case\footnote{Since
$\tr X_1^k$ and $\tr X_2^k$ ( or $\tr \bar{X}_1^k$ and $\tr
\bar{X}_2^k$) with large $k$ becomes relevant at large $x$ (or
$\bar{x}$) one can apply our asymptotic behavior to see where the
operators becomes relevant. }.

%%%%%%%%%%%%%%%%%%%%%%%%%%%%%%%%%%%%%%%%%%%%%%%%%%%%%%%
%%%%%%%%%%%%%%%%%%%%%%%%%%%%%%%%%%%%%%%%%%%%%%%%%%%%%%%%

%%%%%%%%%%%%%%%%%%%%%%%%%%%%%%%%%%%%%%%%%%%%%%%%%%%%%%%%%%%%%%%%%%%%%%%%%%%%%
\section{Higgsing of the E-type Fixed Points}\label{HiggsE}
%%%%%%%%%%%%%%%%%%%%%%%%%%%%%%%%%%%%%%%%%%%%%%%%%%%%%%%%%%%%%%%%%%%%%%%%%%%%%%

In this section we study Higgsing of the $E$-type theories.
In all the $E$-type models the $F$-term condition  demands that the
vev of $Y$  be zero.

\subsection{Higgsing $\widehat{E}$}
We consider Higgsing $U(N)\ra U(N_1)\times U(N_2)$ of the $\widehat{E}$ theory by the
vev eq.(\ref{VEVX}) of the adjoint $X$,
which is allowed by the $F$- and $D$-term conditions.
For the notation for the fields that arise, see
section \ref{HiggsD}.
Bifundamentals coming from
the fluctuations of $X$ become massive while $X_1,X_2,Y_1,Y_2,F,\widetilde{F}$,
$Q_1,Q_2,\widetilde{Q}_1$, and $\widetilde{Q}_2$ remain massless.
In this case, however, at least one of the gauge group factors is asymptotically
non-free and becomes non-interacting in the IR.
When just one factor is asymptotically free,
the analysis of the model proceeds in exactly the same way
as in the $\widehat{A}$ case for the reasons we now explain.
After Higgsing we have the superpotential
\begin{eqnarray}
W=\tr Y_1^3 + \tr Y_2^3 +\tr Y_1F \widetilde{F}+ \tr Y_2
\widetilde{F}F.\label{suM1}
\end{eqnarray}
%Since there is an interaction of singlets with fundamentals it is
%reminiscent of the model studied in \cite{BIWW}, where they
%studied SQCD with singlets.
Assuming that $N_1-N_2>N_f$,
only $SU(N_1)$ is asymptotically free and has $N_f+N_2$ quarks,
two kinds of singlets $Y_2, X_2$ and two adjoints $X_1$ and $Y_1$.
Since the superpotential has $R$ charge two
all the fields that appear in the superpotential have
$R={2\over 3}$.
Thus the gauge singlets can be regarded as free fields as far
as the computation of the trial $a$-function is concerned.
Also we have to consider the anomaly cancellation condition
\begin{eqnarray}
N_1R(X_1)+N_fR(Q_1)-N_f-{N_1+N_2\over 3}=0. \label{i9}
\end{eqnarray}
The trial $a$-function can be written as
\begin{eqnarray}
a=a_{\rm int}(X_1,Q_1)+a(X_2)+a(Y_1)+a(Y_2)+a(Q_1^{\prime}).
\end{eqnarray}
The last four contributions are computed by regarding the singlets
as free fields.
Thus the only difference from $\widehat{A}$ is the constant term
${-N/3}$ in the anomaly cancellation condition eq.(\ref{i9}). Maximizing the trial $a$-function with respect to $R(Q_1)$ yields
\begin{eqnarray}
R(Q_1)=\frac{6 x^2+2 x-y-\sqrt{10 x^4-8 y x^3+2 y^2 x^2-x^2}-3}{3 \left(2 x^2-1\right)}.
\end{eqnarray}
Because $R(Q_1)$ and $R(X_1)$ are monotonically decreasing function of $x,y$ and satisfy $R(Q_1)>{2\over 3}$ and $R(X_1)>{2\over 3}$ for all region $y< x-1$. Therer is thus no reason to doubt the validity of the above experssions. With the central charge of $\widehat{E}$ shown in \cite{IWADE} we verified $a-$theorem conjecture.
%%%%%%%%%%%%%%%%%%%%%%%%%%%%%%%%%%%%%%%%%%%%%%%%%%%%%%%
\begin{figure}[htbp]
\begin{center}
\includegraphics[width=4.0cm,height=3.0cm]
{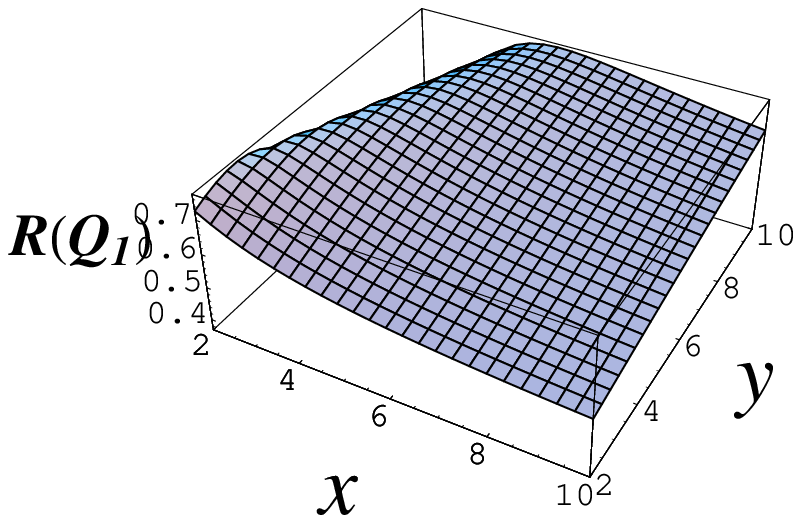}\hspace{1cm}
\includegraphics[width=5.5cm,height=3.0cm]
{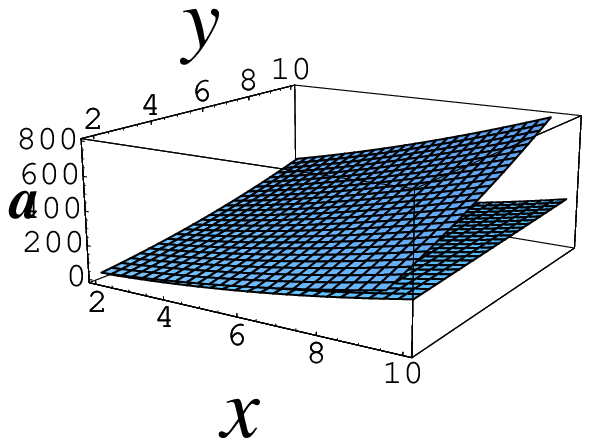}
\end{center}
    \caption{\small The first figure displays $R$ charge $R(Q_1)$ as a function of $x$ and $y$. The second figure shows the central charge of $\widehat{E}$ theory and Higgssed theory. }
    \label{EHiggs}
\end{figure}

\subsection{Higgsing $E_7$}
Let us turn to Higgsing of the $E_k, ~k=6,7,8$ theories.
In the $E_6$ and $E_8$ theories, their superpotentials
do not admit non-zero vevs of $X$ or $Y$,
and there is no Higgsing RG flow allowed.
he superpotential  $\tr (XY^2+Y^3)$ of $E_7$ does
allow us to give a vev to $X$ and admits several breaking patterns:
If we take eq.(\ref{VEVX}) with generic values of $a$ and $b$, the
bifundamentals coming from the fluctuations of $X$ and $Y$, as well
as $Y_1$ and $Y_2$, are massive and get integrated out. Thus the
theory flows to a $SU(N_1)\times SU(N_2)$ theory with massless fields
$X_1,X_2$, $Q_1$, and $Q_2$ and a vanishing superpotential. This is
a product of two copies of $\widehat{A}$. Using the data
for $E_7$ \cite{IWADE} and $\widehat{A}$ \cite{KPS}, we
verified that the $a$-theorem is satisfied under the RG
flow. %\footnote{In calculating the central charge of $E_7$ we
%included unitarity violations of mesonic operators, $\widetilde{Q}Q$
%at $x=6$, $\widetilde{Q}X^iQ$ and $\widetilde{Q}Y^iQ$ at $x=2i+6$
%and $x=3i+6$ with $i=1,2$ respectively. Thus our verification holds
%up to $N/N_f=12$.}
If we take one of vev's in $(\ref{VEVX})$ to be
zero, say $a=0$, $Y_1$ remains massless and there exists a
superpotential $\tr Y_1^3$. However this model is still not
interesting because it is just a product of $\widehat{E}$ and
$\widehat{A}$ without any interactions between them. Again by using
the results in \cite{IWADE} we explicitly checked the $a$-theorem
prediction.

The most interesting possibility is the $a=-b\neq 0$ case. In this case the mass terms
for bifundamentals $F$ and $\widetilde{F}$ coming from $Y$ cancel out. Therefore the
bifundamentals $F$ and $\widetilde{F}$ as well as the adjoint fields
$X_1$ and $X_2$ and the fundamentals $Q_1,Q_2$ remain massless. Also
there are superpotential terms constructed from the bifundamentals:
\begin{eqnarray}
W=\tr (F\widetilde{F})^2+{\rm higher\  order\  terms}.\label{Etp}
\end{eqnarray}
When we integrated out the massive
fields $Y_1$ and $Y_2$ we used the equation of motion.
We drop the higher-order terms,
which are irrelevant when the lowest order term is marginal.
In the rest of the subsection, we focus on this model.

The requirement that the superpotential has $R$ charge two yields
$R(\widetilde{F})=R(F)={1\over 2}$.
The anomaly cancellation conditions are
\begin{eqnarray}
xR(X_1)+R(Q_1)-{y\over 2}=1,\qquad yR(X_2)+R(Q_2)-{x\over 2}=1.
\end{eqnarray}
we are left with two undetermined $R$ charges. To fix them we use
$a$-maximization. The trial $a$-function is
\begin{eqnarray}
a^{(0)}=a[R(X_1),x,y]+a[R(X_2),y,x]+{1\over 8}xy,
\end{eqnarray}
where the last term comes from the contributions of bifundamentals
and we defined the function $a[R,x,y]$ by
\begin{eqnarray}
a[R,x,y]\equiv 2x^2+x^2\left(3(R-1\right)^3-(R-1))+2x
\left(3(xR-{y\over 2})^3-(xR-{y\over 2}) \right).
\end{eqnarray}
Maximizing the trial $a$-function with respect to $R(X_1)$ and
$R(X_2)$, we obtain the $R$ charges of the fields. The expression of
$R(Q_1)$ is relatively simple:
\begin{eqnarray}
R(Q_1)=\frac{12 x^2+6 x-3 y-6-x \sqrt{2 \left(40 x^2-36
   y x+9 y^2-2\right)}}{12 x^2-6}.
\end{eqnarray}
$R(Q_2)$ can be obtained by replacing $x$ with $y$ in $R(Q_1)$.
Using these results it is easy to see that in the region where both
gauge groups are asymptotically free, there are no  gauge invariant
operators that hit the unitarity bound. Therefore our results do not
receive corrections  due to operators hitting the unitarity bound.

\begin{figure}[htbp]
\begin{center}
\includegraphics[width=3.3cm,height=2.5cm]
{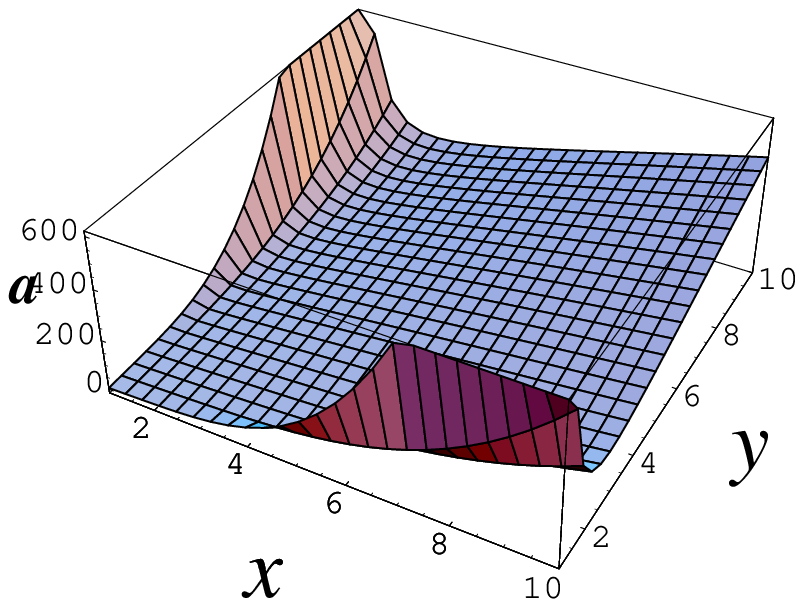}\hspace{1.3cm}
\includegraphics[width=3.5cm,height=2.7cm]
{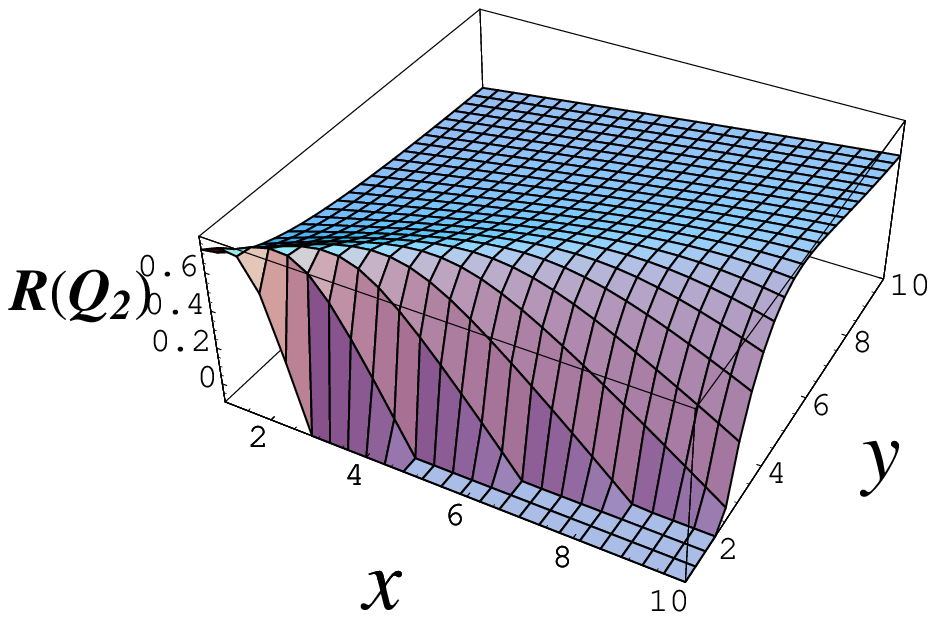}\hspace{1.3cm}
\includegraphics[width=3.0cm,height=2.0cm]
{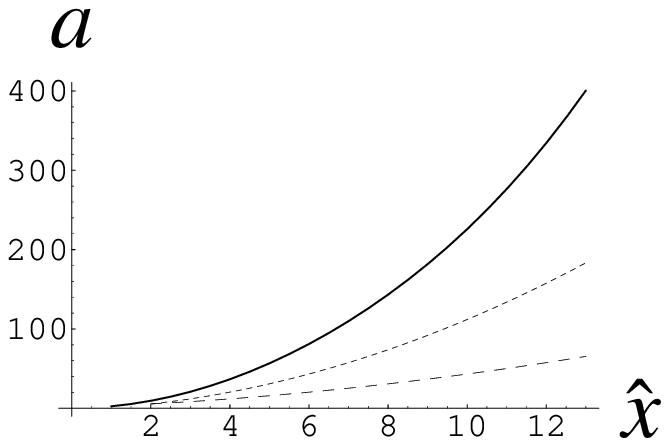}
\end{center}
    \caption{\small The first two figures display the central charge $a$ and the $R$ charge $R(Q_2)$
    as functions of $x$ and $y$. Near the edges of the figures the functions vary abruptly.
    These points are all outside of the region where both gauge groups are
    asymptotically free and are of no significance.
    The third figure shows the central charges of the $E_7$ theory, the Higgsed theories on the  $y=2x-1$
    slice and the $y={x+1\over 2}$ slice from the top to the bottom.
    The $a$-theorem is satisfied.}
    \label{Etypefig}
\end{figure}

The central charge $a(x,y)$ is shown in fig \ref{Etypefig}. With
this function we can check the $a$-theorem conjecture
under the Higgsing RG flow. For simplicity we check the validity of
the conjecture on the lines $y=2x-1$ and $y={x+1\over 2}$, which are
the bounds of asymptotic freedom. In so doing, as in the previous
section,
 we need to be careful about the arguments of the $a$-function.
 For example, when we draw the figure for the central charge on $y=2x-1$,
 we use $a(x,2x-1)|_{x={\widehat{x}+1\over 3}}$, where $\widehat{x} \equiv N/N_f$,
 to compare with the central charge of the $E_7$ theory.
 As we see from figure \ref{Etypefig} the $a$-theorem is satisfied under this Higgsing RG flow.

The region where the Higgsed model is interacting and conformal
may be narrower than that determined by asymptotic freedom of the
gauge groups above,
although we find nothing that suggests this.
If there exists a magnetic description of the $E_7$ theory,
it should be possible to find this region (conformal window)
in a  way similar
to what we did in the $D$-type case.
%%%%%%%%%%%%%%%%%%%%%%%%%%%%%%%%%%%%%%%%%%%%%%%%%%%%%%%
%%%%%%%%%%%%%%%%%%%%%%%%%%%%%%%%%%%%%%%%%%%%%%%%%%%%%%%%

%%%%%%%%%%%%%%%%%%%%%%%%%%%%%%%%%%%%%%%%%%%%%%%%%%%%%%%%%%%%%%%%%%%%%%%%%%%%%
\section{Superpotential Deformations of $\widehat{O}$ }\label{MesonO}
\setcounter{equation}{0}
%%%%%%%%%%%%%%%%%%%%%%%%%%%%%%%%%%%%%%%%%%%%%%%%%%%%%%%%%%%%%%%%%%%%%%%%%%%%%

We start considering  mesonic superpotential deformations. This
section and the following ones can be read independently of the previous ones.
\subsection{Deformations of $\widehat{O}$ by mesonic superpotential terms}

\begin{figure}[htbp]
\begin{center}
\includegraphics[width=10.0cm,height=3.0cm]
{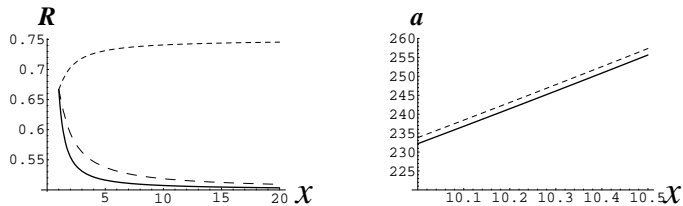}
\end{center}
    \caption{\small The graph on the left displays the $R$ charges $R(Q)$ (top), $R(X)$ (middle),
    and $R(Y)$ (bottom). The graph on the right shows the central charges
    $a_{\widehat{O}}$ (top) and $a_{\widehat{M}}$ (bottom).
    Since the difference is very small the graph is magnified near $x\simeq 10$.
    There is no violation of the  $a$-theorem conjecture anywhere inside or outside of the
    region shown.}
    \label{hat0}
\end{figure}

The authors of \cite{IWADE} studied the RG fixed point, which was
called $\widehat{O}$, of the two-adjoint SQCD with $W=0$ by using
$a$-maximization and calculated the superconformal $U(1)$ $R$
charges and central charge $a$. $R(X)=R(Y)$ and $R(Q)$ are
monotonically decreasing functions of $x=N/N_f$ and take values
 $0.5 \lesssim R(X)< 2/3$ and $0.575 \lesssim R(Q)<2/3$.
Among the operators that preserve
the diagonal $SU(N_f)$,
there are two relevant mesonic superpotential
deformations $\widetilde{Q}_{i}XQ^i\  ({\rm or~ equivalently}\
\widetilde{Q}_{i}YQ^i)$ and $\widetilde{Q}_iQ^i$ for all $x>1$.\footnote{
If we consider deformations by $\sum_{i=1}^k
\widetilde{Q}_i X Q^i$ or $\sum_{i=1}^k \widetilde{Q}_i Q^i$ with
$1\leq k< N_f$, the flavor symmetry is smaller and calculations become
more complicated.  }
 We see that there is no relevant
deformation that includes more than two quark superfields. First let
us consider the deformation by $W=\widetilde{Q}_{i}XQ^i$,
which we expect leads to a new fixed point that we refer to as
$\widehat{M}$. At the fixed point the $R$ charge of $W$ has to be
two, and the $U(1)_R$-$SU(N_i)$-$SU(N_i)$ ABJ anomaly has to vanish.
Thus we have the following two constraints for three independent
charges, leaving one flavor $U(1)$ symmetry at the new fixed
point.\footnote{ Originally the theory has the classical global
symmetry group $SU(N_f)\times SU(N_f)\times U(1)_B \times
U(1)_A\times U(2)_{XY} \times U(1)_{R}$. By adding the mesonic
superpotential terms this classical symmetry group breaks to
$SU(N_f)_{diag}\times U(1)_{diag} \times U(1)_{Y} \times U(1)_{R}$.
Out of the three $U(1)$s we can construct two anomaly free $U(1)$
symmetries.} .
\begin{eqnarray}
2R(Q_i)+R(X)=2,\qquad \ \ \ R(Q_i)+xR(X)+xR(Y)-x-1=0 .\label{Mconst}
\end{eqnarray}
To determine the superconformal $R$ symmetry we use $a$-maximization.
Maximizing the trial $a$-function gives
\begin{eqnarray}
R(X)=\frac{24x^3 - 2x {\sqrt{ -1 - 12x + 98x^2 - 180x^3 + 144x^4 }}}
  {3 - 18x + 30x^2}\label{MRX}.
\end{eqnarray}
Using this result one can check if there is a gauge invariant
operator that hits the unitarity bound. As discussed in \cite{KPS},
${\rm Tr}X^i$ or ${\rm Tr}Y^k$ do not contribute to the central
charge in the large $N$ limit (\ref{largeN}), because the
contributions from these operators are ${\cal O}(N_f^0)$ although
the central charge $a$ is ${\cal O}( N_f^2)$. Generalized mesons and
generalized baryons
\begin{eqnarray}
 {\cal M}_{I_1\cdots I_n}=\widetilde{Q}_i X_{I_1} \cdots X_{I_n}Q_i,
 \qquad B =Q_{(I_1\cdots I_{n_1})}^{n_{(I_1\cdots I_{n_1})}}Q_{(J_1\cdots J_{n_1})}^{n_{(J_1\cdots J_{n_1})}}
 \cdots Q_{(K_1\cdots K_{n_1})}^{n_{(K_1\cdots K_{n_1})}}
\end{eqnarray}
 do contribute to the central charge $a$.
 Here $\left(Q_{(I_1\cdots I_n)}\right)_i \equiv (X_{I_1}\cdots X_{I_{n}}Q)_i$
 and $N=n_{(I_1\cdots I_{n_1})}+\cdots + n_{(K_1\cdots K_{n_1})}$.
 The $R$ charges and the central charge $a$ are shown in figure \ref{hat0}.
From figure \ref{hat0} we see that $R(Q_i)>0.65$ and conclude that
in our new fixed point $\widehat{M}$ there is no meson or baryon
that hits the unitarity bound. Thus  for all $x>1$, our results do
not receive corrections due the decoupling of operators hitting the
unitarity bound. Also we see that the $a$-theorem conjecture holds
under the RG flow.
This is as anticipated because we expect that
the Lagrange multiplier method of \cite{Kutasov} should work in this case.
Still, there is no general proof of the $a$-theorem
for superpotential deformations, and our results are an addition
to the already huge accumulation of evidence supporting the conjecture.
%{\bf This is quite natural from the
%$a$-maximization point of view and proved in \cite{Kutasov,BIWW} by
%using Lagrange multiplier method. Adding a relevant superpotential
%and assuming new RG-fixed point with the superpotential we restrict
%the parameter space of a possible $U(1)_{R}$ symmetry in which trial
%$a$-function can be maximized. Since the parameter space becomes
%small the maximum value should be smaller after the RG-flow. However
%to the best our knowledge there is no general proof for flows
%between CFT points in which there is no ambiguity of $U(1)_R$ and
%thus we can determine the $R$ charges without using
%$a$-maximization. In \cite{Kutasov} the authors studied an examples
%of such kind of flows, namely a flow between $A_{k+1}\to A_{k}$.
%Thus our $a$-theorem verification in the following sections gives a
%non-trivial test. }

Let us consider generalized baryonic operators hitting the unitarity
bound. Since a baryonic operator includes $N$ quark superfields
$Q_i$, in the large $N$ limit  the $R$ charge is of the order
${\mathcal O}(N_f)$. Thus a generalized baryon can hit the unitarity
bound only when $R(Q)$ is very small. Namely $R(Q)=0$ indicates
baryonic operators hitting the unitarity bound\footnote{If one wants
to see the unitarity violation of baryonic operators correctly we
have to evaluate the $R$ charge of the operator $B_{\rm
small}=\prod_{i=0}^{[x]-1}(X_1^iQ_1)^{N_f} $ as did in subsection
2.1.}.
%%%%%%%%%%%%%%%%%%%%%%%%%%%%%%%%%%%%%%%%%%%%%%%%%%%%%%%
%%%%%%%%%%%%%%%%%%%%%%%%%%%%%%%%%%%%%%%%%%%%%%%%%%%%%%%%

%{\bf Since there is no known Seiberg dual description for the
%present gauge theory we expect that there is no stability bound
%which is generated by developing a dynamically generated
%superpotential. One should care the relation between the stability
%bound and an end of the conformal window. As is well known SQCD has
%a conformal window  ${3\over 2}N\le N_f\le 3N$ although in the
%region $N+2 \le N_f \le {3\over 2}$ the theory is a magnetic free
%theory. Since the magnetic free theory is a free CFT we expect that
%$a$-theorem prediction would be satisfied. Therefore when we go
%through an end of conformal window we would not see any $a$-theorem
%violation. However when we reach at $N_f=N+1$ we would see
%$a$-theorem violation if the prediction are true.}

What was found in \cite{IWADE} for their models was that the
following consequences of the $a$-theorem are violated outside the
conformal window: $a(N,N_f)>a(N,N_f-1)$. The inequality comes from
integrating out a quark\footnote{In the paper \cite{IWADE} they
showed the other condition coming from the Higgsing. However in our
case there exists the superpotential term and yields different
theory after Higgsing. Thus it does not hold in our present case.
}. In the large $N$ limit (\ref{largeN}) they can be stated as
\cite{IWADE}
\begin{eqnarray}
{a(x) \over x^2}&& \qquad {\rm must\ be\ a\ monotonically\
decreasing\ function\ of \ }x.\label{lemma}
\end{eqnarray}
Note our convention of $a$ (\ref{aconvention})\footnote{ If a theory
has a stability bound, outside of the bound the theory is not at a
conformal fixed point and
 the $a$-theorem might not hold there.
In fact as demonstrated in \cite{IWADE} the $D_5$ theory, which has
a dual description in terms of a $SU(9N_f-N)$ theory, violates eq.(\ref{lemma}) above the stability bound $N/N_f \simeq 9$. Of course
the violation of the $a$-theorem is not a sufficient condition for
the existence of a stability bound, but it suggests the possibility
that there would be no magnetic dual description.}. We verified that
the conditions in  eq.(\ref{lemma}) are satisfied.
% suggesting the nonexistence of a stability bound in our model.

\begin{figure}[htbp]
\begin{center}
\includegraphics[width=5.5cm,height=3.0cm]
{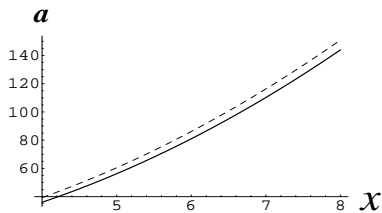}
\end{center}
    \caption{\small Central charge $a$ at $M^{\widehat{M}}_{(0,0)}$ (bottom) and $\widehat{O}$ (top).}
    \label{MQQ}
\end{figure}

Next let us study the mass deformation $W=\widetilde{Q}_{i}Q^i$.
Since quarks are decoupled, we have an asymptotically free gauge
theory with two adjoints. We thus expect the theory to be at a
nontrivial fixed point that we name ${M}^{\widehat{M}}_{(0,0)}$. The
ABJ anomaly cancellation condition can be written as $R(X)+R(Y)=1$.
After $a$-maximization we obtain $R(X)=R(Y)={1\over 2}$ and
$a^{\widehat{M}}_{(0,0)}(x)={9\over 4}x^2$. One can see from figure
\ref{MQQ} that  the $a$-theorem conjecture holds for the flow
$\widehat{O} \rightarrow M^{\widehat{M}}_{(0,0)}$. In this model we can construct exactly marginal operators $\tr X^2Y^2$ and $\tr XYXY$ \cite{LS}. Note that with $R$ charges obtained above we see that $\tr X^4$ is also a marginal operator. However since it breaks the global symmetry that mix the $X$ and $Y$, $\tr X^4$ can not be an exactly margianl operator.

%%%%%%%%%%%%%%%%%%%%%%%%%%%%%%%%%%%%%%%%%%%%%%%%%%%%%%%
%%%%%%%%%%%%%%%%%%%%%%%%%%%%%%%%%%%%%%%%%%%%%%%%%%%%%%%%

In the ${M}^{\widehat{M}}_{(0,0)}$ theory, since quarks have been
integrated out there are no mesons or baryons. As we mentioned in
the $\widehat{M}$ case, operators such as $\tr X^l$  do not
contribute to $a$ in the large $N$ limit. We thus need not consider
the effects of operators hitting the unitarity bound. We checked
that the conditions in eq.(\ref{lemma}) are satisfied for the
${M}^{\widehat{M}}_{(0,0)}$ theory, and we find nothing that
suggests the existence of a stability bound.

There are several relevant deformations $\tr Y^2,\
\tr XY^2$ for all $x>1$ at $M^{\widehat{M}}_{(0,0)}$.
If we assume that these give rise to new fixed points,
we are led to unnatural $R$ charge assignments:
$R(Y)=1 ,\ R(X)=0,\  a(x)=0$.\footnote{
$R(X)=0$ implies that there are many operators that violate the unitarity bound.
Also, $a(x)=0$ is unnatural from the expectation that $a$ counts the massless
degrees of freedom.}
We thus rather conclude that these deformations do not lead to non-trivial fixed points.
 Below, we will meet this type of relevant deformations several times.
 In such a case we
simply ignore them. Finally we  comment on two-adjoint
models realized by D-branes in some Calabi-Yau geometries
\cite{CKV}. These models have superpotentials, constructed from  two
adjoints, which can take the forms we study here. As seen above
among the generic monomial superpotential of the theories only $\tr
Y^2$ and $\tr XY^2$ can be relevant at UV theories. However ABJ
anomaly cancellation condition and marginality of the superpotential
give a vanishing central charge $a(x)=0$. Therefore we conclude that
such models do not have interacting conformal fixed points.

%%%%%%%%%%%%%%%%%%%%%%%%%%%%%%%%%%%%%%%%%%%%%%%%%%%%%%%%%%%%%%%%%%%%%%%%%%%
\subsection{Deformations of $\widehat{M}$ by $\tr X^mY^n$ }
%%%%%%%%%%%%%%%%%%%%%%%%%%%%%%%%%%%%%%%%%%%%%%%%%%%%%%%%%%%%%%%%%%%%%%%%%%%

In this subsection we consider the deformations of $\widehat{M}$ by
$\Delta W=\tr X^mY^n$. Using the $R$ charges (\ref{Mconst}) and
(\ref{MRX}) at $\widehat{M}$, we obtain seven relevant operators,
$\tr X^m Y^n$ with $1<m+n\le 3$ for all range $x>1$. All these
operators have no more than three adjoint fields so there is no
multi-trace operator which give rise to a manifold of fixed points.

\subsubsection{$\widehat{M}\rightarrow M^{\widehat{A}}_{(4,0)}$}
We begin with the adjoint mass term $\tr X^2$. $X$ becomes massive
and gets integrated out. The equation of motion for $X$ is
$2X_{\alpha \beta}+(Q^i\widetilde{Q}_i)_{\alpha \beta}=0$.
Substituting the result into the original superpotential we obtain a
one-adjoint SQCD with $W=-{1\over 2} \widetilde{Q}_j
Q^i\widetilde{Q}_i Q^j$. We call this new fixed point
$M^{\widehat{A}}_{(4,0)}$. Here we see that there exists an operator
that consists of the same fields with different contractions of
flavor indices $\tQ_i Q^i \tQ_j Q^j$. $\widetilde{Q}_j
Q^i\widetilde{Q}_i Q^j$ combined with $\tQ_i Q^i \tQ_j Q^j$ produce
a line of fixed points.

\subsubsection{$\widehat{M}\rightarrow M^{\widehat{D}}_{(0,1)}$}
\begin{figure}[htbp]
\begin{center}
\includegraphics[width=10.0cm,height=3.0cm]
{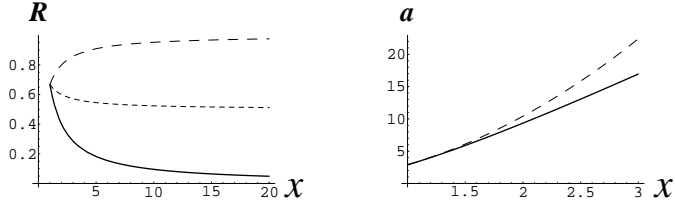}
\end{center}
    \caption{\small The graph on the left shows the $R$ charges $R(X)$ (top), $R(Q)$ (middle) and $R(Y)$ (bottom)
    at the $M^{\widehat{D}}_{(0,1)}$ fixed point. The graph on the right depicts the central charges
    at the fixed points $\widehat{M}$ (top) and $M^{\widehat{D}}_{(0,1)}$ (bottom).}
    \label{tak}
\end{figure}

We now turn to the  $\tr X^2Y$ deformation. Since we have two
superpotential terms, there are two possibilities: 1)We keep the $R$
charges of both terms to be two. 2) We only keep the $R$ charge of
the deformation $\tr X^2Y$ to be two and the other term $
\widetilde{Q}_i X Q^i$ becomes irrelevant under the RG flow. It is worth noting that decreasing of central charge $a$ under the second
assumption is not obvious a priori thought it is clear for the first
one from the $a$-maximization point of view. We computed the $R$
charge of $\tQ_i XQ^i$ under the assumption 2) and found that it is
smaller than two, leading to a contradiction.\footnote{ For examples
appearing later, too, we checked that flows in which the original
superpotential term becomes irrelevant, do not occur.} Thus the
first scenario must happen, where we keep both terms. The flavor
$U(1)$ symmetry is now broken, and  $R$ charges can be determined
without using $a$-maximization: $R(Q)={x+1\over 2x+1}$,
$R(X)={2x\over 2x+1}$, and $R(Y)={2\over 2x+1}$. We name this new
fixed point $M^{\widehat{D}}_{(0,1)}$. We see that $R(Q)>0.5$ and
conclude that there is no unitarity bound violation by operators
including quarks. The central charge is
\begin{eqnarray}
a^{\widehat{D}}_{(0,1)}(x)=\frac{2x^2\left( 2 + 2x + 35x^2 \right)
}{{\left( 1 + 2x \right) }^3}.
\end{eqnarray}
From figure \ref{tak}  we verify that the $a$-theorem holds for the
flow $\widehat{M}\rightarrow M^{\widehat{D}}_{(0,1)}$ . We also
checked that the conditions in eq.(\ref{lemma}) hold.

%%%%%%%%%%%%%%%%%%%%%%%%%%%%%%%%%%%%%%%%%%%%%%%%%%%%%%%
%%%%%%%%%%%%%%%%%%%%%%%%%%%%%%%%%%%%%%%%%%%%%%%%%%%%%%%%

\subsubsection{$\widehat{M}\rightarrow M^{\widehat{E}}_{(1,0)}$ and $\widehat{M}\rightarrow M^{\widehat{E}}_{(0,1)}$}
\begin{figure}[htbp]
\begin{center}
\includegraphics[width=4.5cm,height=2.7cm]
{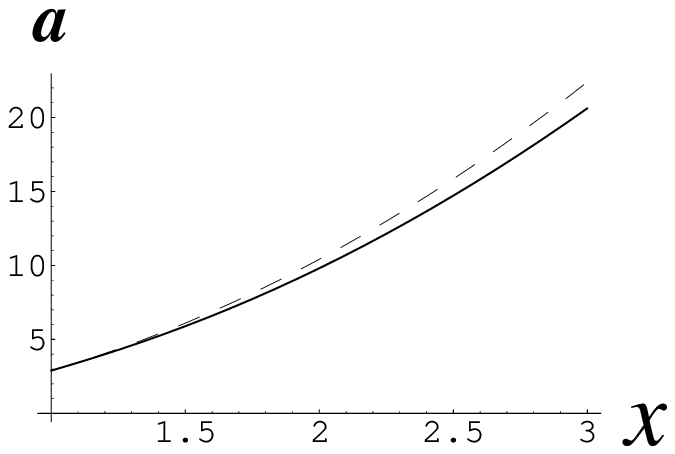}
\includegraphics[width=4.5cm,height=2.7cm]
{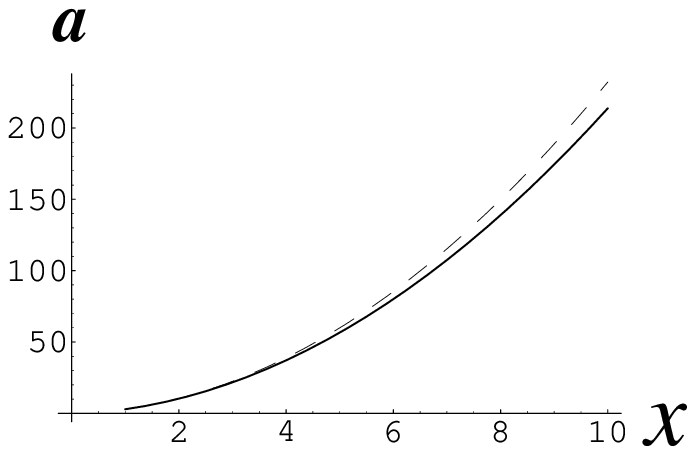}
\end{center}
    \caption{\small
    The solid line on the left graph is $a^{\widehat{E}}_{(1,0)}$ and that on the right graph
    is $a^{\widehat{E}}_{(0,1)}$. The dashed line is  $a_{\widehat{M}}$ in both graphs.}
    \label{atheoremE}
\end{figure}

Next let us study $\tr Y^3$ and $\tr X^3$ deformations. Proceeding
in the same way as in the previous cases we see that these
deformations drive the theory from $\widehat{M}$ to new interacting
CFT points that we name $M^{\widehat{E}}_{(1,0)}$ and
$M^{\widehat{E}}_{(0,1)}$, respectively. The $R$ charges at
$M^{\widehat{E}}_{(1,0)}$ are $R(Q)={5x-3\over 3(2x-1)}$, $R(X)={2x
\over 3(2x-1)}$, and $R(Y)={2\over 3}$. At
$M^{\widehat{E}}_{(0,1)}$, we have $R(Q)={2\over 3}$, $R(X)={2\over
3}$, and $R(Y)={x+1\over 3x}$. In both cases neither baryons nor
mesons hit the unitarity bound. Thus we do not have to consider
operators hitting the unitarity bound. The central charge $a$ can be
written as follows and satisfy the $a$-theorem (figure
\ref{atheoremE}) and eq.(\ref{lemma}):
\begin{eqnarray}
a^{\widehat{E}}_{(1,0)}(x)=\frac{2\,x^2\,\left( 2 + 18\,x - 79\,x^2
+ 72\,x^3 \right) }{9\,{\left( -1 + 2\,x \right) }^3}, \quad
a^{\widehat{E}}_{(0,1)}(x)=\frac{-6x + 1 + 13x^2 + 18x^3}{9x}
\end{eqnarray}
%%%%%%%%%%%%%%%%%%%%%%%%%%%%%%%%%%%%%%%%%%%%%%%%%%%%%%%
%%%%%%%%%%%%%%%%%%%%%%%%%%%%%%%%%%%%%%%%%%%%%%%%%%%%%%%%

\subsubsection{Other deformations}

Finally let us study the remaining deformations, which do not drive
to interacting CFT points. First we consider the deformation $\Delta
W=\tr Y^2$ . Since this is a mass term, $Y$ gets decoupled under the
flow. Thus the theory becomes a one-adjoint SQCD with the
superpotential $\widetilde{Q}_iXQ^i$. We expect that this theory is
not an interacting CFT because $R(X)=0$ and $a(x)=0$ as discussed
the previous subsection.
As for $\Delta W=\tr XY$, integrating out the $X$ and $Y$ leaves a
vanishing superpotential. Therefore we obtain a pure SQCD, for which
the region $x>1$ is out of the conformal window. Likewise using the
ABJ anomaly cancellation condition and marginality of superpotential
$\Delta W=\tr XY^2$ we conclude that it does not drive the theory to
a new interacting CFT point.

%%
%%%%%%%%%%%%%%%%%%%%%%%%%%%%%%%%%%%%%%%%%%%%%%%%%%%%%%%%%%%%%%%%%%%%%%%%%%%%%
%%\subsection{Deformations of $\widehat{M}$ by mesonic superpotential terms}
%%%%%%%%%%%%%%%%%%%%%%%%%%%%%%%%%%%%%%%%%%%%%%%%%%%%%%%%%%%%%%%%%%%%%%%%%%%%%

%
Let us consider further meson deformations of
$\widehat{M}$.
The chiral ring relations $\widetilde{Q}_{i\alpha}
Q^{i\beta}=0$ imply that all the meson operators are
trivial at $\widehat{M}$.
Thus most mesons can be excluded from relevant operators.
The mass deformation $\tQ_i Q^i$ is, however, relevant,
corresponding to the constant shift of $X$.
See footnote \ref{exceptionfoot}.
The theory then flows to $M^{\hat{M}}_{(0,0)}$.

\begin{figure}[htbp]
\begin{center}
\includegraphics[width=6.5cm,height=4.5cm]
{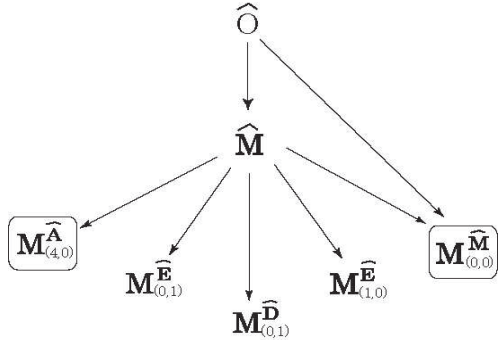}
\end{center}
    \caption{\small New interacting superconformal fixed points and flows. The squar means that there exists an exactly margianl operator in all region of the RG-fixed point.}
    \label{FlowFromO}
\end{figure}

In this section we have considered several flows from the
$\widehat{O}$ and obtained new interacting fixed points. These flows
and fixed points are summarized in figure \ref{FlowFromO}. As we
will see later all these new fixed points also arise by mesonic
superpotential deformations of $\widehat{A}$, $\widehat{D}$, and
$\widehat{E}$.

%%%%%%%%%%%%%%%%%%%%%%%%%%%%%%%%%%%%%%%%%%%%%%%%%%%%%%%
%%%%%%%%%%%%%%%%%%%%%%%%%%%%%%%%%%%%%%%%%%%%%%%%%%%%%%%

%%%%%%%%%%%%%%%%%%%%%%%%%%%%%%%%%%%%%%%%%%%%%%%%%%%%%%%%%%%%%%%%%%%%%%%%%%%%%
\section{Superpotential Deformations of $\widehat{E}$}\label{MesonE}
\setcounter{equation}{0}
%%%%%%%%%%%%%%%%%%%%%%%%%%%%%%%%%%%%%%%%%%%%%%%%%%%%%%%%%%%%%%%%%%%%%%%%%%%%%

\subsection{Deformations of $\widehat{E}$  by mesonic superpotential terms}

In this section we study mesonic superpotential deformations of the
fixed point $\widehat{E}$, which is characterized by $W=\tr Y^3$.
The authors of  \cite{IWADE} discussed deformations of the form
$\Delta W={\rm Tr} X^m Y^n$. We extend their study to deformations
by mesonic terms. From the data for $\widehat{E}$ in table
\ref{hatEresult}, taken from \cite{IWADE}, and by taking the chiral
ring relation $Y^2=0$ into account, we see that there are eight
mesonic terms that become relevant for some values of $x$: $\Delta
W=  \widetilde{Q}_iXYQ^i,\widetilde{Q}_iYXQ^i,
\widetilde{Q}_iYQ^i,\tQ_j Q^i\tQ_i Q^j$, and $\widetilde{Q}_iX^lQ^i$
with $0\le l\le 3$ \footnote{In addition to these operators we have different kind of relevant operators, $\tr XY \tQ_i Q^i$, $\tr X^2 \tQ_i Q^i$, $\tr X^2 \tQ_i XQ^i$ and $\tr X^3 \tQ_i Q^i$. First two are the exactly marginal operators at $M^{\widehat{E}}_{(1,1)}, M^{\widehat{E}}_{(2,0)}$ and the last two are those at $M^{\widehat{E}}_{(3,0)}$. Note that if we assume $U(N)$ gauge theory we have more exactly marginal operators. }.
%({\bf We need to be careful about the mesons hitting the unitarity bound.})
The symbols for these deformations were defined in
eq.(\ref{MhatEsymbols}).
In the regions where these operators are
relevant, see table \ref{hatEresult}. We have to pay attention to
the order of $X$ and $Y$. $\widetilde{Q}_iXYQ^i$ and
$\widetilde{Q}_iYXQ^i$ are distinct operators, and we expect that
they give rise to a line of fixed points by the argument given in
the introduction. Also $\tQ_j Q^j \tQ_i Q^i$ combines with the fourth
operator to yields a line of fixed points.
The anomaly cancelation and the marginality of  $W={\rm Tr}Y^3+\Delta W$
is enough to determine the $R$
charges $R(Q_i)$, $R(Y)$ and $R(X)$ without using
$a$-maximization\footnote{ It is a priori possible
that ${\rm Tr}Y^3$ becomes irrelevant after the deformation by $\Delta W$.
We explicitly checked by using $a$-maximization that this does not occur. }.
From the $R$ charges, one can determine the range of $x$ where
$\Delta W$ is relevant.
We checked that there is no operator hitting the unitarity bound
in this range for the flow from $\widehat{E}$ to each fixed point.
These results are summarized in table \ref{hatEresult}.
 One can verify from the sample values of $a$ given in the table
 that all the flows satisfy the conjectural $a$-theorem.
 To be complete, we list the $a$-functions at the fixed points.
 \ba
a_{(3,0)}^{\widehat{E}}={2x^2(8x^3-29x^2+18x+6)\over(2x-3)^3}&,&
a_{(2,0)}^{\widehat{E}}={x^2(18x^3-41x^2+18x+4)\over 9(x-1)^3},\nn\\
a_{(1,0)}^{\widehat{E}}={2x^2(72x^3-79x^2+18x+2)\over 9(2x-1)^3}&,&
a_{(0,1)}^{\widehat{E}}={18x^3+13x^2-6x+1 \over 9x},\nn\\
a_{(1,1)}^{\widehat{E}}={2x^2(8x^3-5x^2-10x+6)\over (2x-1)^3} &,&
a_{(0,0)}^{\hat{E}}=2x^2,\label{aE}\\
a^{\widehat{E}}_{(4,0,0)}(x)=- \frac{3}{2}   + \frac{3}{8\,x} +
\frac{7\,x}{4} + 2\, x^2 .\nn
 \ea

Note that at $M^{\widehat{E}}_{(0,0)}$
$R(X)$ is independent of $x$.
One can check that the beta function for the gauge coupling and the beta functions for $\tr X^4Y$ and $\tr X^6$
are linearly dependent, giving rise to a line of fixed points.

%%%%%%%%%%%%%%%%%%%%%%%%%%%%%%%%%%%%%%%%%%%%%%%%%%%%%%%%%%%%%%%%
\begin{table}[htbp]
\begin{center}
\begin{tabular}{|c||c|c|c|c|c|c|}
\hline
  &$W$&  {\small $R(X)$}  & {\small $R(Q)\equiv y$}  & relevant & {\footnotesize $a(50)$} & {\footnotesize $a(30)$}
  \\  \hline \hline
 $\widehat{E}$ &$\Tr Y^3$ &   ${1+x-y\over x}-{2\over3}$ &  ${1+{x(2-\sqrt{10x^2-1})\over 3(2x^2-1)}}$  & $x>1$ &{\footnotesize $5086.20$} & {\footnotesize $1851.08$ }\\ \hline
${ M}_{(3,0)}^{\hat{E}}$ & $\Tr Y^3+\tQ_i X^3 Q^i$&  ${2x \over 3(-3+2x)}$ &    ${x-3\over
2x-3}$  &  {\small $ x\ge 40.8$} &{\footnotesize
$5086.19$} & {\footnotesize $1851.06$ } \\ \hline ${
M}_{(2,0)}^{\hat{E}}$  &$\Tr Y^3+\tQ_i X^2 Q^i$& ${x\over 3(-1+x)}$ & ${2x-3\over 3(x-1)}$
 & {\small $x\ge 3.7$}& {\footnotesize $5072.54$}
& {\footnotesize $1843.63$ } \\ \hline
 ${ M}_{(1,0)}^{\hat{E}}$&$\Tr Y^3+\tQ_i X Q^i$ &  ${2x\over 3(-1+2x)}$ &  ${5x-3\over 3(2x-1)}$  & $x>1$ &
 {\footnotesize $5040.49$} & {\footnotesize $1824.38$ } \\ \hline
 ${ M}_{(0,1)}^{\hat{E}}$ &$\Tr Y^3+\tQ_i Y Q^i$ & ${1+x\over 3x}$ & ${2\over 3}$& $x>1$& {\footnotesize $5071.56$} &
 {\footnotesize $1842.67$ } \\ \hline
 ${ M}_{(1,1)}^{\hat{E}}$ & $\Tr Y^3+\tQ_i X YQ^i$& ${2(1+x)\over 3(-1+2x)}$  & ${x-1\over 2x-1}$ & {\small $x\ge 21.3$}
  & {\footnotesize $5086.09$} & {\footnotesize $1851.06$ } \\ \hline
 ${ M}_{(0,0)}^{\hat{E}}$  &$\Tr Y^3+\tQ_i Q^i$& ${1\over 3}$ &Integrated out &$x>1$& {\footnotesize $5000.00$} & {\footnotesize $1800.00$ } \\ \hline
 ${ M}_{(4,0,0)}^{\hat{E}}$  &$\Tr Y^3+(\tQ_i Q^i)(\tQ_i Q^i)$& ${1\over 3}+{1\over 2x}$ & ${1\over 2}$  &{\small $x\ge 11.6$ }& {\footnotesize $5086.01$} & {\footnotesize $1851.01$ } \\ \hline
\end{tabular}
\end{center}
\caption{
Data for mesonic term deformations of $\widehat{E}$.
$R(Y)$ is always $2/3$ and is omitted.
The range of $x$ where there is a flow from $\widehat{E}$ to each
fixed point is indicated.
Since drawing graphs of the $a$-function for all the flows is tedious, we list
sample values of $a$ to test the $a$-theorem.} \label{hatEresult}
\end{table}
%%%%%%%%%%%%%%%%%%%%%%%%%%%%%%%%%%%%%%%%%%%%%%%%%%%%%

%%%%%%%%%%%%%%%%%%%%%%%%%%%%%%%%%%%%%%%%%%%%%%%%%%%%%%%%%%%%%%%%%%%%%%%%%%%%%
\subsection{Deformations of $M^{\widehat{E}}_\ast$  by mesonic superpotential terms \label{meson52}}
%%%%%%%%%%%%%%%%%%%%%%%%%%%%%%%%%%%%%%%%%%%%%%%%%%%%%%%%%%%%%%%%%%%%%%%%%%%%%%

Next we consider flows between the fixed points
found in the previous subsection.

We consider deforming $W_0+\Delta W=\Tr Y^3+\Delta W$ in the previous subsection
further by a relevant operator $\Delta W'$.
There are a priori four possibilities after the deformation:
1) $W_0$ and $\Delta W$ become irrelevant, leaving only $\Delta W'$.
2) $\Delta W$ become irrelevant, and $W_0$ and $\Delta W'$ remain.
3) $W_0$ become irrelevant, and $\Delta W$ and $\Delta W'$ remain.
4) $W_0$ and $\Delta W'$ become irrelevant and $\Delta W$ remain.
It turns out that 1) never occurs.\footnote{
We explicitly checked in all examples that
the $R$ charges determined by $a$-maximization
are inconsistent with the assumption 1).}
4), where the deforming operator becomes irrelevant,
seems unlikely and we do not consider this possibility.
({ It is of course better to explicitly exclude
this possibility by $a$-maximization.})
For most deformations, the $R$ charges obtained by assuming 3)
is inconsistent with the assumption, and 2) is what actually occurs.
There is just one deformation, for which we cannot conclusively exclude 3).
This is discussed in appendix \ref{exception}.
Even in this case, we argue that 3) is what  happens.

%And then we study the
%deformation by  operators $\Delta W^{\prime}$. In doing so we
%have two possibilities, because we already have two superpotential
%terms, $\tr Y^3+\Delta W$. One is to keep the $R$ charge of $\tr
%Y^3$ to be two, and the other is to keep the one of $\Delta W$ to be
%two. If we keep the $\tr Y^3$, then $\Delta W$ must be irrelevant.
%If it is not irrelevant, we conclude that this flow does not
%occur

%Most of the calculation need in this
%subsections were already done in previous section. We can simply use
%them and check at the fixed point the original superpotential is a
%irrelevant or not.
%Because the $R$ charges of fields are positive,
%there are always flows where a higher order term is replaced
%by a lower order term:
%\begin{eqnarray}
%M^{\widehat{E}}_{(3,0)}\to M^{\widehat{E}}_{(2,0)}\to
%M^{\widehat{E}}_{(1,0)}\to M^{\widehat{E}}_{(0,0)},\quad
%M^{\widehat{E}}_{(1,1)} \to M^{\widehat{E}}_{(0,1)} \to
%M^{\widehat{E}}_{(0,0)},\ \ {\rm and} \ \   M^{\widehat{E}}_{(0,2)}
%\to M^{\widehat{E}}_{(0,1)}. \nonu
%\end{eqnarray}
%Because all the $R$ charge of $X$ and $Y$ at
%$M^{\widehat{E}}_{(m,n)}$ are positive, if we keep the $R$ charge of
%$\tQ XQ$ to be two, then that of $\tQ X^lQ$ must be larger than two,
%namely irrelevant. Now we are ready to study deformation from each
%$M^{\widehat{E}}_{(m,n)}$.

\subsubsection{$M^{\widehat{E}}_{(3,0)}\ra M^{\widehat{E}}_{(2,0)}$,
$M^{\widehat{E}}_{(3,0)}\ra M^{\widehat{E}}_{(1,1)}$,
$M^{\widehat{E}}_{(3,0)}\ra M^{\widehat{E}}_{(0,0)}$,
$M^{\widehat{E}}_{(3,0)}\ra M^{\widehat{E}}_{(1,0)}$,
$M^{\widehat{E}}_{(3,0)}\ra M^{\widehat{E}}_{(0,1)}$, and
$M^{\widehat{E}}_{(3,0)}\ra M^{\widehat{E}}_{(4,0,0)}$}
Let us begin by considering mesonic deformations of
$M^{\widehat{E}}_{(3,0)}$ in the range $x\geq 40.8$, where
the fixed point $M^{\widehat{E}}_{(3,0)}$ exists.
Using the results in table 1 we obtain
the following relevant operators at this fixed point:
\begin{eqnarray}
\widetilde{Q}_{i}X^2Q^i,~
[\widetilde{Q}_{i}XYQ^i,\widetilde{Q}_{i}YXQ^i],~
\widetilde{Q}_{i}Q^i ,~
\widetilde{Q}_{i}XQ^i,~
\widetilde{Q}_{i}YQ^i,~{\rm and}~~
[(\widetilde{Q}_jQ^i)(\widetilde{Q}_iQ^j),(\widetilde{Q}_iQ^i)(\widetilde{Q}_jQ^j)]. \label{relE30}
\end{eqnarray}
The paired operators give rise to lines of fixed points
as discussed in the introduction.

%The relevant operator $\widetilde{Q}_iX^2Q^i$ in
%(\ref{relE30}) is trivial in the chiral ring
%by virtue of the third relation in eq.
%(\ref{chiralringEm0}).
%Nevertheless we  should include
%the operator in discussing RG flows because the relation
%$\widetilde{Q}_iX^2Q^i\simeq 0$ comes from a constant shift of $X$,
%in other words, from $X$ acquiring an expectation value.
%This is a real physical effect, unlike a generic field redefinition.
%It moves the theory  to a different super-selection sector.
%The shift of $X$ can be interpreted as fluctuation
%around an expectation value, $\langle X \rangle+\delta X$. Therefore
%if we include the flows by a Higgsing, this flow does occur.
%({\bf I removed the discussion of chiral ring relations because
%we actually do not use them here at all.})

As an example of the flow from $M^{\widehat{E}}_{(3,0)}$ let us
consider the deformation by $\tQ XYQ$.
In this case one can easily check that 2) occurs.
The values of the central charge at the original and
final final points, for sample values $50$ and $30$ of $x$, are
$(a_{(3,0)}^{\widehat{E}}(50),a_{(3,0)}^{\widehat{E}}(30))=(5086.19,1851.06)$
and
$(a_{(1,1)}^{\widehat{E}}(50),a_{(1,1)}^{\widehat{E}}(30))=(5086.09,1851.06)$
respectively.
We see that $a$ decreases under the flow.
We checked that the $a$-theorem conjecture holds for all values of $x$.

The deformation by
$\widetilde{Q}_jQ^i\widetilde{Q}_iQ^j$ is subtle and is discussed in detail
in appendix \ref{exception}.
We argue that 2) is what occurs, yielding a flow $M^{\widehat{E}}_{(3,0)} \ra M^{\widehat{E}}_{(4,0,0)}$.

\subsubsection{
$M^{\widehat{E}}_{(2,0)}\ra M^{\widehat{E}}_{(1,0)}$,
$M^{\widehat{E}}_{(2,0)}\ra M^{\widehat{E}}_{(0,1)}$,
$M^{\widehat{E}}_{(2,0)}\ra M^{\widehat{E}}_{(0,0)}$,
$M^{\widehat{E}}_{(1,0)}\ra M^{\widehat{E}}_{(0,0)}$,
$M^{\widehat{E}}_{(0,1)}\ra M^{\widehat{E}}_{(1,0)}$, and
$M^{\widehat{E}}_{(0,1)}\ra M^{\widehat{E}}_{(0,0)}$
}
In the same way we consider mesonic deformation from
$M^{\widehat{E}}_{(2,0)}$, $M^{\widehat{E}}_{(1,0)}$ and
$M^{\widehat{E}}_{(0,1)}$.
The relevant operators
at $M^{\widehat{E}}_{(2,0)}$ are $ \widetilde{Q}_{i}XQ^i,~ \widetilde{Q}_{i}YQ^i$,
and $\widetilde{Q}_{i}Q^i$.
These operators yield RG flows whose end points are
$M^{\widehat{E}}_{(1,0)}$, $M^{\widehat{E}}_{(0,1)}$, and $M^{\widehat{E}}_{(0,0)}$,
respectively. Using the $R$ charges shown in table \ref{hatEresult}
we can check that the original operator $\tQ_i X^2 Q^i$ becomes
irrelevant at the end points of the RG flows. At
$M^{\widehat{E}}_{(1,0)}$ there is only one relevant operator
$\tQ_i Q^i$ that produces a flow from $M^{\widehat{E}}_{(1,0)}$ to $M^{\widehat{E}}_{(0,0)}$.
As for $M^{\widehat{E}}_{(0,1)}$ there
are two relevant operators $\tQ_i XQ^i$ and $\tQ_i Q^i$.
One can check that these operators drive the theory to
$M^{\widehat{E}}_{(1,0)}$ and $M^{\widehat{E}}_{(0,0)}$,
respectively.

\subsubsection{$M^{\widehat{E}}_{(1,1)}\ra M^{\widehat{E}}_{(1,0)}$,
$M^{\widehat{E}}_{(1,1)}\ra M^{\widehat{E}}_{(0,1)}$,
$M^{\widehat{E}}_{(1,1)}\ra M^{\widehat{E}}_{(2,0)}$,
$M^{\widehat{E}}_{(1,1)}\ra M^{\widehat{E}}_{(0,0)}$, and
$M^{\widehat{E}}_{(1,1)}\ra M^{\widehat{E}}_{(4,0,0)}$
}
At $M^{\widehat{E}}_{(1,1)}$,
relevant operators are
\begin{eqnarray}
\widetilde{Q}_{i}XQ^i, ~\widetilde{Q}_{i}YQ^i,~
\widetilde{Q}_{i}X^2Q^i,~ \widetilde{Q}_{i}Q^i,~{\rm and}~
[(\tQ_jQ^i)(\tQ_iQ^j), (\tQ_iQ^i)(\tQ_jQ^j)].
\end{eqnarray}
%Again we neglect the relevant operator only in the region where
%accidental symmetry appears.
For the first four among the relevant
operators, we can proceed in the same way as previous cases, thus we
skip the details. On the other hand last one is quite different from
the others. Let us study the deformation by $\tQ_j Q^i\tQ_i Q^j$
operator. If we keep  $R(\tQ_iXYQ^i)$ to be two, we see that there
is no solution compatible to ABJ anomaly condition. On the other
hand, if we keep the $R$ charge of $\tr Y^3$ to be two we obtain
$R(Y)={2\over 3}$, $R(X)=\frac{1}{3}+\frac{1}{2x}$, and $R(Q)=1/2$.
This is exactly the same as $M^{\widehat{E}}_{(4,0,0)}$
which are already discussed earlier. Using this $R$ charge we see
that $\tQ_iXYQ^i$ is irrelevant under the flow. We explicitly
checked $a$-theorem and (\ref{lemma}) by using (\ref{aE}).

\subsubsection{$M^{\widehat{E}}_{(4,0,0)}\ra M^{\widehat{E}}_{(2,0)}$,
$M^{\widehat{E}}_{(4,0,0)}\ra M^{\widehat{E}}_{(1,0)}$,
$M^{\widehat{E}}_{(4,0,0)}\ra M^{\widehat{E}}_{(0,1)}$, and
$M^{\widehat{E}}_{(4,0,0)}\ra M^{\widehat{E}}_{(0,0)}$}
%Finally we comment on the mesonic deformation of
%$M^{\widehat{E}}_{(4,0,0)}$. Using the chiral ring relation we
%obtain three equations, $(Y^2)^\alpha_{~\beta}= (\tQ_k Q^i)\tQ_i=Q^i
%(\tQ_i Q^k)=0$. From the last two expressions we obtain
%$(Q^i\tQ_i)^\alpha_{~ \beta}=0$ and conclude that there is no mesonic
%deformations.
On the line of fixed points $M^{\widehat{E}}_{(4,0,0)}$,
there are four relevant operators that produce flows
to $M^{\widehat{E}}_{(2,0)}, M^{\widehat{E}}_{(1,0)},
M^{\widehat{E}}_{(0,1)}$, and $M^{\widehat{E}}_{(0,0)}$,
respectively.

%%%%%%%%%%%%%%%%%%%%%%%%%%%%%%%%%%%%%%%%%%%%%%%%%%%%%%%
\begin{figure}[htbp]
\begin{center}
\includegraphics[width=8.0cm,height=7.0cm]
{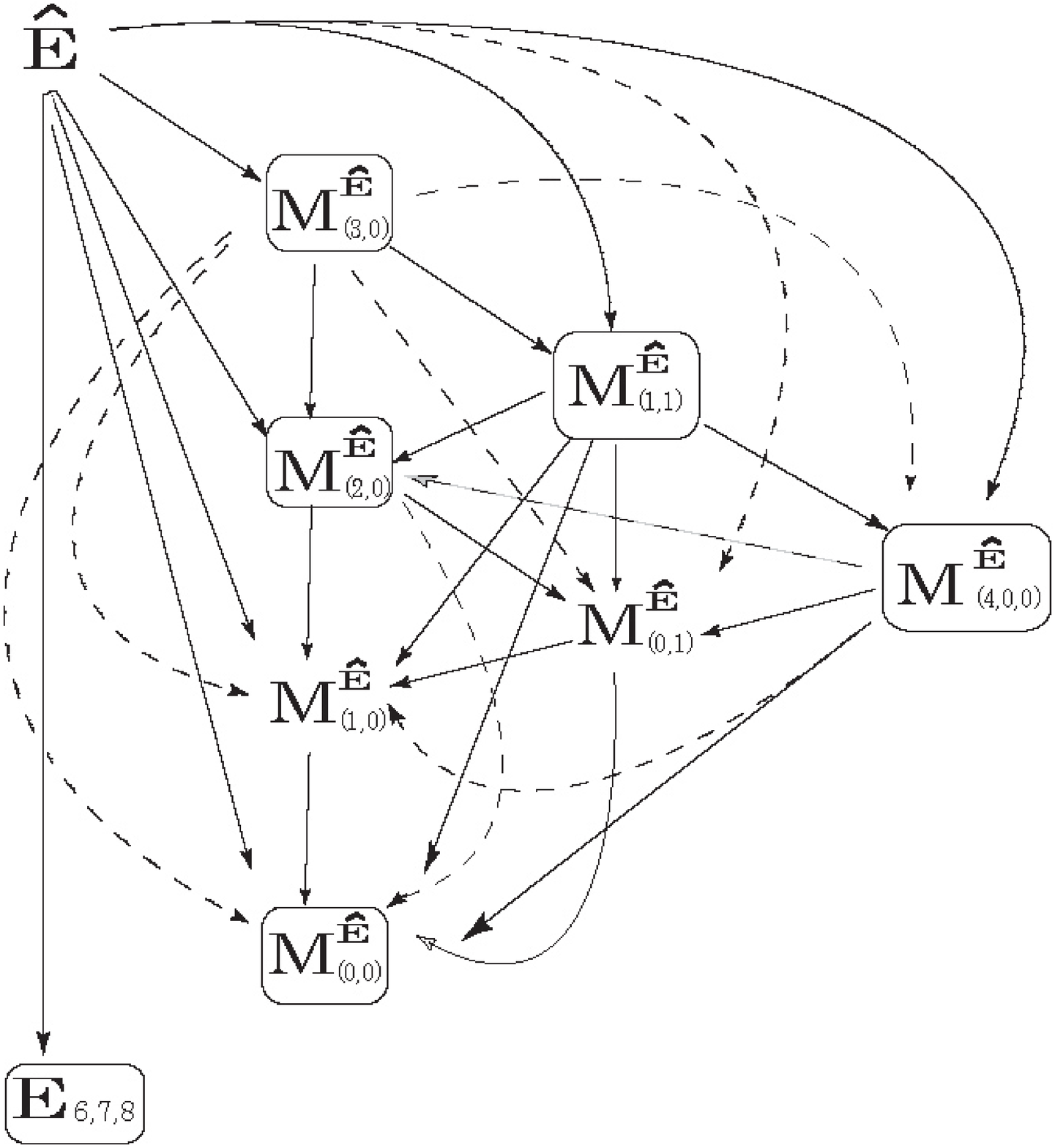}
\end{center}
    \caption{\small New fixed points obtained from $\widehat{E}$ by RG flows
    caused by meson operators.
    The arrows indicate allowed RG flows.
    Dashed lines are used just to make the relations easier to understand.}
    \label{FlowFromE}
\end{figure}
%%%%%%%%%%%%%%%%%%%%%%%%%%%%%%%%%%%%%%%%%%%%%%%%%%%%%%%%

In this subsection we have explored flows caused by meson operators,
starting from $\widehat{E}$.
The new fixed points and the flows between them
 are summarized in figure
\ref{FlowFromE}.
%These meson type fixed points are ``intermediate
%states'' between $\widehat{E}$ and $E_6$ $E_7$ and $E_8$.
%So
%deformation by $\tr X^mY^n$ type superpotential would drive to $E_6$
%$E_7$ and $E_8$. In the next subsection we will consider this issue.

%%%%%%%%%%%%%%%%%%%%%%%%%%%%%%%%%%%%%%%%%%%%%%%%%%%%%%%%%%%%%%%%%%%%%%%%%%%%%%%
\subsection{Deformations of $M^{\widehat{E}}_\ast$ by $ \tr X^mY^n$ }
%%%%%%%%%%%%%%%%%%%%%%%%%%%%%%%%%%%%%%%%%%%%%%%%%%%%%%%%%%%%%%%%%%%%%%%%%%%%%%%

Take $M^{\widehat{E}}_{(3,0)}$ as an example.
Relevant operators of the form $\Delta W'=\Tr X^mY^n$ at the fixed point
$M^{\widehat{E}}_{(3,0)}$, which has $W_0+\Delta W=\Tr Y^3+\tQ_i X^3 Q^i$, are
\begin{eqnarray}
&&\tr X^5, \tr X^3Y, \tr X^4,\tr X^2Y, \tr X Y, \tr X^2,~
{\rm and}~ \tr Y^2,\label{opeXY}
\end{eqnarray}
where we listed only inequivalent deformations,
using the chiral ring relation $Y^2\sim 0$ and the
fact that the deformation by $\tr X^3$ is equivalent to the deformation by
$\tr X^2Y$ by a change of variables.
The first four deformations do not lead to
an interacting fixed point, since
none of the assumptions 1)-3) at the beginning of this subsection
is consistent in the range of $x$
where $M^{\widehat{E}}_{(3,0)}$ exists.
The remaining deformations $\tr X^2$, $\tr Y^2$ and $\tr XY$ are mass
deformations.
The deformation $\tr Y^2$ makes $Y$ massive, leading to the
one-adjoint SQCD with the superpotential $\tQ_i X^3Q^i$.
This theory cannot be an interacting CFT because
the assumption $R(\tQ_i X^3Q^i)=2$ is inconsistent with unitarity.
The deformation $\tr XY$ makes both $X$ and $Y$ massive
and leads to SQCD that does not have a stable vacuum
in the range $x>1$.
The $\tr X^2$ deformation deforms
$M^{\widehat{E}}_{(3,0)}$ to the $A_2$ fixed point.
%\footnote{Actually, when solve
%the equation of motion for $X$ we have one other solution, $X=2/3
%(Q\widetilde{Q})^{-1}$. However matrix $Q\tQ$ does not have inverse
%matrix so that is not a solution.}

Relevant $\Tr X^m Y^n$ deformations of other fixed points of the form $M^{\widehat{E}}_\ast$ can be
 studied similarly.
Most of them do not lead to an interacting fixed point.
The exception is the mass deformation by $\Tr X^2$ of $M^{\widehat{E}}_{(m,0)},~m=1,2,3$ that
leads to the $A_2$ theory.
\section{Superpotential Deformations of $\widehat{A}$ \label{MesonA}}
%%%%%%%%%%%%%%%%%%%%%%%%%%%%%%%%%%%%%%%%%%%%%%%%%%%%%%%%%%%%%%%%%%%%%%%

Deformations by $\Tr X^{k+1},~k=2,3,...$ of the $\widehat{A}$
fixed point were considered
in \cite{KPS} and lead to the fixed points $A_k$.
In this section, we consider deformations by other types of
superpotential terms.

The $\hat{A}$ theory is asymptotically free for $x>1/2$.
The $R$ charges are obtained by $a$-maximization,
taking into account the corrections due to the mesons hitting the unitarity bound.
$R(Q)=:y(x)$ and $R(X)=(1-y(x))/x$ are monotonically decreasing
functions of $x$ with asymptotics $R(Q)\sim 1-\frac{\sqrt{5}}{3}\simeq 0.244$
and $R(X)\sim \frac{4-\sqrt{3}}{3x} $
as $x\ra \infty$ \cite{KPS}.

We are interested in relevant deformations constructed from mesons
$\Mcal_l\equiv \tQ X^{l-1} Q$.
Since no meson can have $R$ charge smaller than $2/3$,
we consider  a single meson
or a product of two mesons: $\tQ_i X^{l-1} Q^i$,
$(\tQ_i X^{l_1-1} Q^j)(\tQ_i X^{l_2-1} Q^j)$
 and $(\tQ_i X^{l_1-1} Q^i)(\tQ_j X^{l_2-1} Q^j)$.
More than two mesons cannot give a relevant operator.

The deformation of $\hat{A}$ by
$\tQ_i X^{l-1} Q^i$ does not lead to an interacting fixed point
for $x>1/2$ as we now see.\footnote{At the boundary
$x=1/2$, the theory with $l=2$ flows to the $\Ncal=2$ QCD with $N_f=2N_c$
that is conformal.}
If $R(\tQ_i X^{l-1} Q^i)>2/3$ at $\widehat{A}$,
the assumption that $R(\tQ_i X^{l-1} Q^i)=2$ at the potential new fixed point
leads to the $R$ charges $R(X)=0$ and $R(Q)=1$ that are inconsistent
with unitarity.
If $\tQ_i X^{l-1} Q^i$ hits the unitarity bound at $\hat{A}$
and is decoupled,
the term linear in a free field cannot give
an interacting CFT.
By the same argument we conclude that an operator of the type
$\tr X^a (\tQ_i X^b Q^i)$ does not lead to a new CFT.

\begin{figure}[htbp]
\begin{center}
\includegraphics[width=4.0cm,height=2.5cm]
{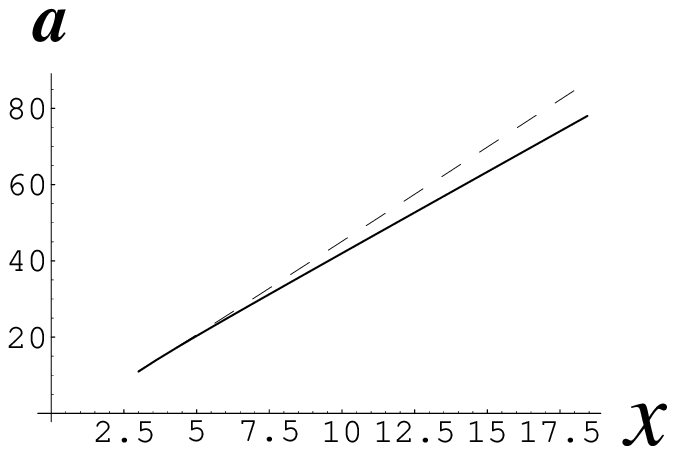}\hspace{1cm}
\includegraphics[width=4.0cm,height=2.5cm]
{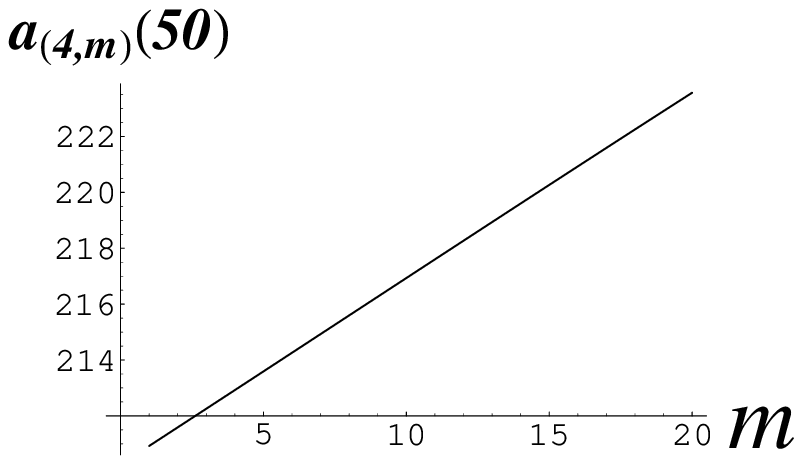}
\end{center}
    \caption{\small Left: The central charges
    at $\hat{A}$ (dashed) and $M^{\widehat{A}}_{(4,3)}$ (solid).
    Right: The central charge at $M^{\hat{A}}_{(4,m)}$
    at $x=50$. It decreases under the flow $M^{\hat{A}}_{(4,m)} \ra M^{\hat{A}}_{(4,m-1)}$.}
        \label{Aflowfig}
\end{figure}

We fix $m \geq 0$ and consider deformations by $(\tQ_i X^{l_1-1} Q^j)(\tQ_i X^{l_2-1} Q^j)$
 and $(\tQ_i X^{l_1-1} Q^i)(\tQ_j X^{l_2-1} Q^j)$ with $l_1+l_2-2=m$ altogether.
Also we restrict ourselves to the deformations that do not contain free mesons.
These deformations give rise to a manifold of fixed points,
which we refer to as $M^{\hat{A}}_{(4,m)}$,
as discussed in the introduction.
The $R$ charges at $M^{\hat{A}}_{(4,m)}$ are
$R(X)={2\over 4x-m}$ and $R(Q)={2x-m\over 4x-m}$.
Both of them are positive  in the range where the deformations
are relevant.
%({\bf I checked this. It follows from $y(x)>0$.})
Some mesons
violate the unitarity bound and get decoupled.
If we denote by $a_0$ the central charge computed
without the decoupled operators taken in to account,
the correct central charge is modified to
\begin{eqnarray}
a^{\widehat{A}}_{(4,m)}=a_0+{1\over 9}\sum_{l=1}^{\infty}
\left[2-3R({\cal M}_l) \right]^2\left[5-3R({\cal M}_l) \right]
\theta \left(m-{3\over 2}(l-1)-x \right),
\end{eqnarray}
where $\theta$ is the Heaviside step function.
We checked that the $a$-theorem conjecture is satisfied
under the flows $\hat{A}\ra M^{\widehat{A}}_{(4,m)}$
and $M^{\widehat{A}}_{(4,m)}\ra M^{\widehat{A}}_{(4,m-1)}.$
See figure \ref{Aflowfig}.
%%%%%%%%%%%%%%%%%%%%%%%%%%%%%%%%%%%%%%%%%%%%%%%%%%%%%%%%

\begin{figure}[htbp]
\begin{center}
\includegraphics[width=2.0cm,height=4.0cm]
{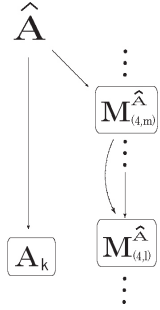}
\end{center}
    \caption{\small Flows from $\widehat{A}$ by mesonic deformations.
    There is no flow that goes between $A_k$ and
    the theory $M^{\widehat{A}}_{(4,m)}$ obtained by a mesonic deformation of $\widehat{A}$.}
    \label{FlowsA}
\end{figure}

The new fixed points and the flows between them are summarized in
figure \ref{FlowsA}.
%({\bf Figure \ref{FlowsA} should be corrected.})
%%%%%%%%%%%%%%%%%%%%%%%%%%%%%%%%%%%%%%%%%%%%%%%%%%%%%%%
%%%%%%%%%%%%%%%%%%%%%%%%%%%%%%%%%%%%%%%%%%%%%%%%%%%%%%%%

%%%%%%%%%%%%%%%%%%%%%%%%%%%%%%%%%%%%%%%%%%%%%%%%%%%%%%%%%%%%%%%%%%%%%%%
\section{Superpotential Deformations of $\widehat{D}$\label{MesonD} }
%%%%%%%%%%%%%%%%%%%%%%%%%%%%%%%%%%%%%%%%%%%%%%%%%%%%%%%%%%%%%%%%%%%%%%%

In this section, we consider mesonic deformations of $\widehat{D}$.
Procedures we will use is the same as the ones in previous sections.
Also the basic features are the similar to $A$-type, so we will
simply show the results.

Using the chiral ring relations $\{X,Y\}=Y^2=0$ and
the $R$ charges shown in \cite{IWADE} we obtain the
following relevant mesonic deformations at $\widehat{D}$,
\begin{eqnarray}
\left[ \tQ_i X^{k} Q^i,~(\tr X^a)\tQ_i X^b Q^i \right],~\left[(\tQ_i X^{l_1} Q^i)(\tQ_j X^{l_2} Q^j),~
(\tQ_i X^{l_1} Q^j)(\tQ_j X^{l_2} Q^i)\right],\\
\left[\tQ_iX^{l}YQ^i,~(\tr X^a)\tQ_i X^b Y Q^i ,\quad (\tr X^aY)\tQ_i X^b Q^i \right],\\
\left[(\tQ_i X^{l_1} Q^i)(\tQ_j X^{l_2} YQ^j),~
(\tQ_i X^{l_1} Q^j)(\tQ_j X^{l_2} YQ^i)\right],
(\tr X^aY)\tQ_i X^b Y Q^i.
\end{eqnarray}
Operators in $[...]$ give a single manifold of fixed points.

%{\bf In addition to these operators we have another kind of relevant operators,}\begin{eqnarray}
% ,\quad ,\quad (\tr X^aY)\tQ_i X^b Y Q^i,
%\end{eqnarray}
%{\bf with $a+b=k$. First one drive to the same CFT points as $\tQ_i X^k Q^i$ and second and third ones are the exactly marginal operators of $\tQ_i X^k Y Q^i $. However the last one drive to new CFT points.}

The deformation of $\hat{D}$ by $\tQ_i X^{l} Q^i$
does not lead to a new fixed point.
This is because requiring that $\Tr X Y^2$ and $\tQ_i X^{l} Q^i$ have
$R$ charge 2,
we get $R(Q)=1, R(X)=0, R(Y)=1$,
which are inconsistent with unitarity.
%({\bf I think there is a possibility that
%all $\Tr X^l$ that violate the unitarity bound
%get decoupled and that the central charge is modified to
%be a positive value.})

\vspace{0.3cm}
\noindent
{\bf Deformations of $\hat{D}$ by $(\tQ_i X^{l_1} Q^i)(\tQ_j X^{l_2} Q^j)$ and $(\tQ_i X^{l_1} Q^j)(\tQ_j X^{l_2} Q^i)$}
\vspace{0.3cm}

\begin{figure}[htbp]
\begin{center}
\includegraphics[width=4.5cm,height=3.0cm]
{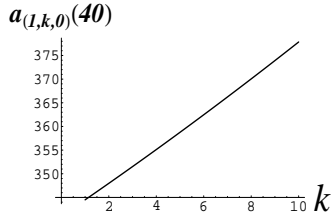}\hspace{1.5cm}
\end{center}
    \caption{\small The central charge at $M^{\widehat{A}}_{(1,k,0)}(40)$.
    % and $a^{\widehat{D}}_{(m,1)}(50)$
    }
    \label{Datho}
\end{figure}

The operators $(\tQ_i X^{l_1} Q^i)(\tQ_j X^{l_2} Q^j)$ and
$(\tQ_i X^{l_1} Q^j)(\tQ_j X^{l_2} Q^i)$ with $l_1+l_2=k$
produce a manifold of fixed points
that we call $M^{\hat{D}}_{(1,m,0)}$.
The $R$ charges are
$R(Q)=\frac{x-k}{2x-k}, R(X)=\frac{2}{2x-k}, R(Y)=\frac{2x-k-1}{2x-k}$.
We checked for $k \le 5$ and $k\gg 1$ that $R(X)$ is positive in the region where the operator
$(\tQ X^{l_1}Q)(\tQ X^{l_2}Q)$ are relevant.
%({\bf We should check that $R(X)$ is positive.})
There are flows $M^{\hat{D}}_{(1,k,0)}\ra M^{\hat{D}}_{(1,k-1,0)}$.
We checked that the $a$-theorem is satisfied under this flow.
See figure \ref{Datho}.

%%%%%%%%%%%%%%%%%%%%%%%%%%%%%%%%%%%%%%%%%%%%%%%%%%%%%%%
%%%%%%%%%%%%%%%%%%%%%%%%%%%%%%%%%%%%%%%%

\vspace{0.3cm}
\noindent
{\bf Deformations of $\hat{D}$ by $\tQ_i X^{l}Y Q^i$}
\vspace{0.3cm}

The deformation of $\hat{D}$ by $\tQ_i X^{l}Y Q^i$
leads to a fixed point that we call
$M^{\hat{D}}_{(0,l,1)}$.
$X$ has a positive $R$ charge \footnote{ We checked for $k \le 5$ and $k\gg 1$ that $R(X)$ is positive in the region where the operator $(\tQ X^{k}YQ)$ are relevant. },
and there is a flow
$M^{\hat{D}}_{(0,l,1)}\ra M^{\hat{D}}_{(0,l-1,1)}$.
We checked that the $a$-theorem is satisfied here. Since the operators $(\tr X^a)\tQ_i X^b YQ^i$ and $(\tr X^a Y)\tQ_i X^b Q^i$ with $a+b=k$ give the same beta function we regard these operators as the exactly marginal opetrators.

\vspace{0.3cm}
\noindent
{\bf Deformations of $\hat{D}$ by $(\tQ_i X^{l_1} Q^i)(\tQ_j X^{l_2} YQ^j)$
and $(\tQ_i X^{l_1} Q^j)(\tQ_j X^{l_2} YQ^i)$}
\vspace{0.3cm}

$(\tQ_i X^{l_1} Q^i)(\tQ_j X^{l_2} YQ^j)$
and $(\tQ_i X^{l_1} Q^j)(\tQ_j X^{l_2} YQ^i)$ with $l_1+l_2=k$
for fixed $k$ give rise to a manifold of fixed points that
we call $M^{\hat{A}}_{(1,k,1)}$.
The central charge is
\begin{eqnarray}
a^{\widehat{D}}_{(k,1)}(x)=\frac{2\,x^2\, \left( 2 + 152\,k^2 + 2\,x
+ 35\,x^2 - 4\,k\,\left( 1 + 38\,x \right)  \right) }
  {{\left( 1 - 4\,k + 2\,x \right) }^3}.
\end{eqnarray}
Again we checked the positivity of $R(X)$. Thus there is a flow
$M^{\hat{D}}_{(1,k,1)}\ra M^{\hat{D}}_{(1,k-1,1)}$.
We checked that the $a$-theorem is satisfied here.

\vspace{0.3cm}
\noindent
{\bf Deformations of $\hat{D}$ by $(\tr X^a Y)\tQ_j X^{b} YQ^j$}
\vspace{0.3cm}

Likewise $R$ charges are given as $R(X)={2\over x-k+1}$, $R(Y)={x-k \over x-k+1 }$ and $R(Q)={1-k \over x-k+1}$. We checked positivity of $R(X)$ for $k\gg 1$ in the region where $(\tr X^a Y)\tQ_j X^{b} YQ^j$ is relevant and $M^{\widehat{D}}_{(0,k,2)}$ $\to $  $M^{\widehat{D}}_{(0,k,2)}$. We checked a-theorem for several model with $k\le 4$.

\vspace{1cm}

\begin{figure}[htbp]
\begin{center}
\includegraphics[width=6.0cm,height=6.0cm]
{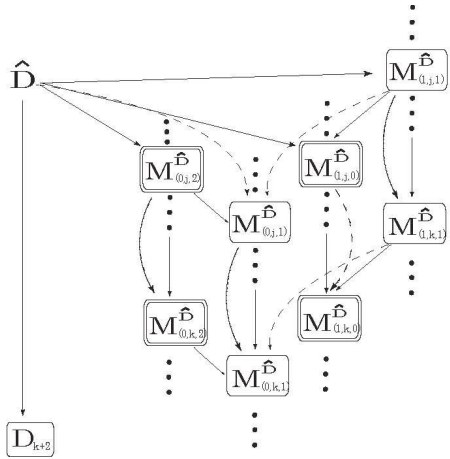}
\end{center}
    \caption{\small Flows driven by mesonic deformation of $\widehat{D}$.}
    \label{Dflow}
\end{figure}

Finally let us comment on deformations of
$M^{\widehat{D}}_{(0,m,1)}$  by  $\Tr X^mY^n$. Using the $R$ charges shown above we
see there are several relevant operators of this type if we take
$x$ to be large enough. However from the chiral ring relation some
of operators are equivalent to certain mesonic deformations and only
one type of deformations becomes independent, $\tr X^k$ with
${4m+3k-1\over 2}>x>{4m+k-1\over 2}$. If the superpotential $QX^mYQ$
is irrelevant under the flow, it drive the theory to $D_{k+1}$.
Unfortunately in this case there appear infinite number of baryonic
operators which hit the unitarity bound. Therefore we could not
explicitly check if this flow occurs or not. From the results of $E$
and $A$-type results we expect that there is no flow between
$D_{k+2}$ with mesonic type fixed points. All the RG flows driven by
mesonic deformation are summarized in figure \ref{Dflow}.

%%%%%%%%%%%%%%%%%%%%%%%%%%%%%%%%%%%%%%%%%%%%%%%%%%%%%%%
%%%%%%%%%%%%%%%%%%%%%%%%%%%%%%%%%%%%%%%%%%%%%%%%%%%%%%%%

%%%%%%%%%%%%%%%%%%%%%%%%%%%%%%%%%%%%%%%%%%%%%%%%%%%%%%%%%%%%%%%%%%%%%%%%%%%%%%
\section{Summary and Discussion}\label{Summary}
%%%%%%%%%%%%%%%%%%%%%%%%%%%%%%%%%%%%%%%%%%%%%%%%%%%%%%%%%%%%%%%%%%%%%%%%%%%%%%

In this paper, we studied large classes of RG flows, triggered by
either scalar expectation values or superpotential terms.

Sections \ref{HiggsD} and \ref{HiggsE} were devoted to
Higgsing of
the $D$- and $E$-type RG fixed points.
As an example of Higgsing of a $D$-type theory, in section \ref{HiggsD} we took the flow
from $\hat{D}$
to the model in \cite{Brodie}, which breaks the gauge group as
$U(N)\to U(N_1)\times U(N_2)$ and ends up with a superpotential
constructed from bifundamental as well as adjoint fields.
By exploiting the existence of a Seiberg duality for this model,
we proposed a prescription, generalizing similar ones used
in other cases, which determines
the two-dimensional conformal window of the model, which is parameterized by
$(x,y)=(N_1/N_f,N_2/N_f)$.
We focused on the cases $k=3$ and $k\gg 1$,
and determined the values of $(x,y)$,
in the large $N$ limit eq.(\ref{largeN}),
at the corners of the conformal window.
This allowed us do draw a picture of the two-dimensional
conformal window in each of the $k=3$ and $k\gg 1$ cases (figure \ref{CWindow}).
In section \ref{HiggsE} we considered Higgsing of $E$-type theories and studied
the non-Abelian Coulomb phase of the model after  Higgsing and
checked the validity of the $a$-theorem.
Compared with the RG flows triggered by superpotential deformations
or gauge interactions,
we still have little understanding of how the $a$-theorem works
for Higgsing RG flows.
Still, we have added more to the evidence supporting
the conjectural $a$-theorem for the Higgsing RG flows.

In sections \ref{MesonO}-\ref{MesonD}, we studied mesonic and monomial superpotential
deformations of $ADE$-type fixed points. We explored a large number of
new RG-fixed points and the RG-flows between them. All of the new
RG-fixed points can be reached from $\widehat{O}$ by
some superpotential deformations.  We did not see any violation
of the $a$-theorem.
Also we pointed out that when we classify relevant operators
by using the chiral ring relations,
we need to consider a constant shift of a field separately though it is an example
of a field redefinition.

We extended in many directions the classes of conformal fixed points
obtained from a two-adjoint gauge theory with flavors.
Our hope was that this would give a clue to understand
the appearance of the $ADE$ classification, as explained in
the introduction.
Unfortunately, the mystery remains unsolved,
and we feel that it is an interesting future direction.

%%%%%%%%%%%%%%%%%%%%%%%%%%%%%%%%%%%%%%%%%%%%%%%%%%%%%%%%%%%%%%%%%%%%%%%%%%%%%%%%%%%%%%%%%%%%%%%%%%%%%%%%%%%%%%%%%%%%%%%%%%%%%%%%%%%%%%%%%%%%%%%%%%%%%%%%%%%%%%%%%%%%%%%%%%%%%%%%%%%%%%%%%%%%%%%%%%%%%%%%%%%%%%%%%%%%%%%%%%%%%%%%%%%%%%%%%%%%%%%%

%%%%%%%%%%%%%%%%%%%%%%%%%%%%%%%%%%%%%%%%%

\section*{Acknowledgments}
\indent Y.O. would like to thank Tohru Eguchi, Teruhiko Kawano, Yu
Nakayama and Yuji Tachikawa for discussions. Y.O. also thanks the
California Institute of Technology for hospitality, where part of
this work was done. T.O. was supported in part by DOE grant
DE-FG03-92-ER40701. Y.O. was supported by a JSPS Research
Fellowship.

%%%%%%%%%%%%%%%%%%%%%%%%%%%%%%%%%%%%%%%%%%%%%%%%%%%%%%%%%%%%%%%%%%%%%%%%%%%%%
\appendix

\renewcommand{\theequation}{\Alph{section}\mbox{.}\arabic{equation}}

\bigskip\bigskip
\noindent {\LARGE \bf Appendix}

%%%%%%%%%%%%%%%%%%%%%%%%%%%%%%%%%%%%%%%%%%%%%%%%%%%%%%%%%%%%%%%%%%%%%%%%%%%
\section{Perturbative Calculation}
%%%%%%%%%%%%%%%%%%%%%%%%%%%%%%%%%%%%%%%%%%%%%%%%%%%%%%%%%%%%%%%%%%%%%%%%%%%

In this appendix we will check the existence of the RG fixed points $\widehat{M}$,
$M^{\widehat{D}}_{(0,1)}$, $M^{\widehat{E}}_{10}$ and $M^{\widehat{E}}_{(0,1)}$ by
perturbative calculations. Since we are interested in the
large $N,N_f$ limit with fixed $x=N/N_f$, the perturbative region can be represented
by $x\simeq 1+\epsilon$ with $0\le \epsilon \ll 1$.
The beta functions of gauge coupling and Yukawa interactions are listed
in \cite{MV} up to two-loop order in perturbation. At one-loop order the superpotential deformations do not affect the gauge coupling beta function and as shown in \cite{IWADE} gauge coupling at the fixed point can be written as
\begin{eqnarray}
{g_\ast^2N_c\over 8\pi^2}\simeq {\epsilon \over 4}.
\end{eqnarray}
Below we will use show the beta functions of the RG fixed points $\widehat{M}$, $M^{\widehat{D}}_{(0,1)}$, $M^{\widehat{E}}_{10}$ and $M^{\widehat{E}}_{(0,1)}$ and check the existence.

\vspace{0.3cm}
\noindent
{\bf  $ W=\lambda \tQ X Q$ at the $\widehat{M}$ RG fixed point }
\vspace{0.3cm}

From the formula $(2.10)$ in \cite{MV} we obtain the anomalous dimensions
of the fields $Q$, $\tQ$ and $X$ as follows:
\begin{eqnarray}
\gamma (Q)&=&\gamma (\tQ)={\lambda^2 \over 64\pi^2}N -{1\over 16\pi^2}g^2 N \nonu \\
\gamma (X)&=& {\lambda^2 N \over 64\pi^2}-{1\over 8\pi^2}g^2N
\end{eqnarray}
Plugging these into the beta function, we obtain
\begin{eqnarray}
\beta (\lambda)= {1\over 2}\lambda  \left[\gamma (X)+2\gamma (Q) \right]={1\over 2}\lambda \left[ 3{\lambda^2 \over 64\pi^2}N -{\epsilon \over 2} \right] .
\end{eqnarray}
At the new RG fixed point, the coupling constant becomes
\begin{eqnarray}
\lambda^2_{\ast}={32\pi^2 \over 3N} \epsilon.
\end{eqnarray}

\vspace{0.3cm}
\noindent
{\bf  $ W=\lambda_1 \tQ X Q+{1\over 6}\lambda_2 \tr X^3$ at the $M^{\widehat{E}}_{(0,1)}$ RG fixed point }
\vspace{0.3cm}

In the same way we examine $M^{\widehat{E}}_{(0,1)}$.
The beta functions for the two coupling constants in the superpotential are
\begin{eqnarray}
\beta (\lambda_1)&=&{1\over 2}\lambda_1 (\gamma(X)+2\gamma (Q))=
 {\lambda_1\over 2 }\left[ {3\lambda_1^2N+2\lambda_2^2d^2 \over 64\pi^2}-{\epsilon \over 2}\right]  \nonu \\
\beta (\lambda_2)&=&{3\over 2}\lambda_2 \gamma (X)={3\over 2}\lambda_2\left[{\lambda_1^2N+2\lambda_2^2d^2 \over 64\pi^2}-{\epsilon \over 4} \right].
\end{eqnarray}
Again the contribution of superpotential is negative and drive the theory to new RG fixed points where the coupling constants have the following values,
\begin{eqnarray}
{\lambda_1^2}_{\ast}={8\pi^2 \epsilon \over N},\quad {\lambda_2^2}_{\ast}={4\pi^2 \epsilon \over d^2}.
\end{eqnarray}
Here $d^2$ is defined by
\begin{eqnarray}
d_{abc}d^{ebc}=d^2 \delta_{a}^{e},\qquad {\rm where} \quad \ d^{abc}=\tr [ T^a\{T^bT^c\}].
\end{eqnarray}

\vspace{0.3cm}
\noindent
{\bf  $ W=\lambda_1 \tQ X Q+{1\over 6}\lambda_2 \tr Y^3$ at the $M^{\widehat{E}}_{(1,0)}$ RG fixed point }
\vspace{0.3cm}

We proceed to the calculation of beta function of $M^{\widehat{E}}_{(1,0)}$.
In this case $\gamma (Q)$ and $\gamma (X)$ are the same as the ones for $\widehat{M}$ and althought $\gamma (Y)$ is diffrent.
\begin{eqnarray}
\beta (\lambda_1)={1\over 2}\lambda_1 \left[ {3\lambda^2_1 \over 64\pi^2}N -{\epsilon \over 2} \right],\quad  \beta (\lambda_2)={2\over 3}\lambda_2 \left[{\lambda_2^2 d^2 \over 32\pi^2}-{g^2N \over 8\pi^2} \right]
\end{eqnarray}
\begin{eqnarray}
{\lambda^2_1}_{\ast}={32\pi^2 \over 3N} \epsilon,\quad {\lambda_2^2}_{\ast}={8\pi^2  \over d^2}\epsilon.
\end{eqnarray}

\vspace{0.3cm}
\noindent
{\bf  $W=\lambda_1 \tQ X Q+{1\over 2}\lambda_2 \tr X^2Y$ at the $M^{\widehat{D}}_{(0,1)}$ RG fixed point}
\vspace{0.3cm}

Last one is $M^{\widehat{D}}_{(0,1)}$. In this case $\gamma (Q)$ is the same as the one for $\widehat{M}$.
Plugging it into the formula for the beta function, we obtain
\begin{eqnarray}
\beta (\lambda_1)={\lambda_1\over 2} \left[{3\lambda_1^2N+4\lambda_2^2d^2 \over 64\pi^2}-{\epsilon \over 2} \right], \qquad
\beta (\lambda_2)= {\lambda_2\over 2}\left[{\lambda_1^2N+5\lambda_2^2d^2 \over 32\pi^2}-{3\epsilon \over 4} \right].
\end{eqnarray}
At the new RG fixed point the coupling constants have values
\begin{eqnarray}
{\lambda^2_1}_{\ast}={64\pi^2 \over 11N} \epsilon,\quad {\lambda_2^2}_{\ast}={40\pi^2  \over 11 d^2}\epsilon.
\end{eqnarray}

%%%%%%%%%%%%%%%%%%%%%%%%%%%%%%%%%%%%%%%%%%%%%%%%%%%%%%%%%%%%%%%%%%%%%%%%%%%%%
\section{Higgsing of $A$-type Fixed Points} \label{HiggsA}
We give a brief review of Higgsing of the $\widehat{A}$ theory studied
in \cite{KPS}. When the adjoint field $X$ acquires a vacuum
expectation value,
\begin{eqnarray}
\langle X \rangle ={\rm
diag}(\stackrel{N_0}{\overbrace{0,\cdots,0}},
\stackrel{N_1}{\overbrace{a_1,\cdots,a_1}},\cdots
\stackrel{N_n}{\overbrace{a_n,\cdots,a_n}}),\qquad \sum_{i=0}^nN_i
=N  \label{AtypeVEV}
\end{eqnarray}
the gauge group $SU(N)$ breaks into a product
group $\prod SU(N_i)$. Since bifundamental matter fields coming from
the fluctuations of the original $X$ are eaten, the $SU(N_i)$
factors do not interact with each other. Thus the theories become a
product of several $\widehat{A}$ theories and do not provide new interacting CFT points.

Adding a mesonic superpotential $W=\widetilde{Q}X^mQ$ and
considering Higgsing, can we obtain different kinds of non-trivial
fixed points? Plugging (\ref{AtypeVEV}) back into the superpotential we obtain the following superpotentials,
\begin{eqnarray}
W= \widetilde{Q}_0 X_0^m Q_0 +\sum_{i=1}^n\sum_{k=0}^m (a_i)^k
\widetilde{Q}_iX_i^{m-k}Q_i,
\end{eqnarray}
where we decomposed the original quark superfields $Q$ into $n+1$
subsectors $Q_i$, and $X_i$ is an adjoint superfield of $SU(N_i)$
group. We omitted the flavor index of $Q_i$. In the subsectors with a nonzero vev, quark superfields
$Q_i$ with $i=1,\cdots n$ are massive and thus are integrated out in
the IR, although $Q_0$, being in the subsector with a vanishing vev,
remains massless. The
marginality of the first superpotential term gives $2R(Q_0)+mR(X_0)=2$ and the ABJ anomaly cancellation condition can be written as
$N_0R(X_0)+N_fR(Q_0)=N_f$. Combining these two conditions we conclude
that $R(X_0)=0$, $R(Q_0)=1$ and $a=0$. As in similar cases in the text, we conclude that
there is no non-trivial fixed point.

What if we further add $\tr X^{k+1}$, namely Higgsing of the
$A_{k+1}$ model? Again we assume the vev (\ref{AtypeVEV}), after
Higgsing, the adjoint superfields of $SU(N_i)$ with $i=1,\cdots n$ have mass
terms and are integrated out. Thus those subsectors are driven to the
product of several copies of SQCD that does not have any non-trivial
fixed point in the range $N/N_f>1$. On the other hand a subsector
$SU(N_0)$ has a superpotential term $\tr X_0^k$ which is the
$A_{k+1}$ theory. Thus we conclude that Higgsing of the $A$-type
theories (i.e., $\widehat{A}$ and $A_{k+1}$) does not lead to new
interacting fixed points. This is in contrast with the $D$-type
($\widehat{D}$ and $D_{k+2}$) and $E$-type ($\widehat{E}$ and
$E_{6,7,8}$) theories where there are several breaking patterns
which drive the theory to new interacting fixed points.

\section{Undetermined Flow: $(\widetilde{Q}Q)^2$ deformation of $M_{(3,0)}^{\widehat{E}}$ \label{exception}}

Among the many RG flows we studied, there was one case where we could
not determine which flow actually occurs. That is the deformation of
$M^{\widehat{E}}_{(3,0)}$ by $\Delta W^{\prime}
=\widetilde{Q}_jQ^i\widetilde{Q}_iQ^j$. The operator becomes
relevant for $x>{15\over 4}$ although $M^{\widehat{E}}_{(3,0)}$
exists only in $x>40.8$. Hence  we focus on the region where both
conditions are satisfied.

First we study the case in which $R$ charges of $\tr Y^3$ and
$\widetilde{Q}_jQ^i\widetilde{Q}_iQ^j$ are two, assuming that $\widetilde{Q}_iX^3Q^i$ is irrelevant under the RG-flow.
$R$ charge at the fixed point can be given as $R(Y)={2\over 3}$,
$R(X)={1\over 3}+{1\over 2x}$, and $R(Q)={1\over 2}$. There is no
operator which violates the unitarity bound. With these $R$ charges let us
check if operator $\tQ_i X^3Q^i$ is irrelevant or not:
$R(\tQ_iX^3Q^i)={2}+{3\over 2x} >2$. Thus there is not
contradiction to our assumption. The $a$-theorem also satisfied in
this flow.

Let us next consider the another possibility and
assume that $\tr Y^3$ is irrelevant after the RG flow. In this case $R$
charges are given by $ R(X)={1\over 3}$, $R(Y)={2\over 3}+{1\over
2x}$ and $R(Q)={1\over 2}$. We call this new fixed point
$M^{(3,0)}_{(4,0,0)}$. $\tr Y^3$ is irrelevant after the flow: $R(\tr Y^3)=2+{3\over 2x}>2$.
The central charge $a$ in this fixed point also
satisfies the $a$-theorem.

Which is the real flow\footnote{ Using
$a$-maximization we explicitly checked one of the $R$ charges of the superpotential terms is less than two
if we make the third assumption that the two terms in the original superpotential
become irrelevant. Thus the assumption is not correct and can be excluded. }? We expect that the latter one would actually occurs. As we see in the
main text we obtain the RG-flows $M^{\widehat{E}}_{(3,0)} \to
M^{\widehat{E}}_{(1,1)}$ and $M^{\widehat{E}}_{(1,1)}\to
M^{\widehat{E}}_{(4,0,0)}$. So combining these two flows we can obtain an
indirect flow from $M^{\widehat{E}}_{(3,0)}$ to
$M^{\widehat{E}}_{(4,0,0)}$.
On the other hand we can not reach $M^{(3,0)}_{(4,0,0)}$ from
another RG-fixed points obtained in the main text. Therefore it is plausible  that the RG flow $M^{\widehat{E}}_{(3,0)} \to
M^{\widehat{E}}_{(1,1)}$ occurs.

%%%%%%%%%%%%%%%%%%%%%%%%%%%%%%%%%%%%%%%%%%%%%%%%%%%%%%%%%%%%%%%%%%%%%%%%%%%%

\end{document}